\documentclass{emulateapj}

\usepackage{color}

\newcommand{\simgt}{\lower.5ex\hbox{$\; \buildrel > \over \sim \;$}}
\newcommand{\simlt}{\lower.5ex\hbox{$\; \buildrel < \over \sim \;$}}

\def\ls{\mathrel{\hbox{\rlap{\hbox{\lower4pt\hbox{$\sim$}}}\hbox{$<$}}}}
\def\gs{\mathrel{\hbox{\rlap{\hbox{\lower4pt\hbox{$\sim$}}}\hbox{$>$}}}}

\def\V606{\mathrel{V_{606}}}

\slugcomment{Zhang et al. 2010, ApJ, 711, 1033-1043}

\shorttitle{LoCuSS: A Comparison of Cluster Mass Measurements from
  \emph{XMM-Newton} and Subaru} 
\shortauthors{Zhang, Okabe,
  Finoguenov, et al.}

\begin{document}


\title{LoCuSS: A Comparison of Cluster Mass Measurements from
  \emph{XMM-Newton} and Subaru --- Testing Deviation from Hydrostatic
  Equilibrium and Non-Thermal Pressure Support\altaffilmark{*}}

\altaffiltext{*}{This work is based on observations made with the
  \emph{XMM-Newton}, an ESA science mission with instruments and
  contributions directly funded by ESA member states and the USA
  (NASA), and data collected at Subaru Telescope and obtained from the
  SMOKA, which is operated by the Astronomy Data Center, National
  Astronomical Observatory of Japan.}


\author{Yu-Ying Zhang\altaffilmark{1}} 
\email{yyzhang@astro.uni-bonn.de}

\author{Nobuhiro Okabe\altaffilmark{2,3}}

\author{Alexis Finoguenov\altaffilmark{4,5}}

\author{Graham P. Smith\altaffilmark{6}}

\author{Rocco Piffaretti\altaffilmark{7}}

\author{Riccardo Valdarnini\altaffilmark{8}}

\author{Arif Babul\altaffilmark{9}}

\author{August E. Evrard\altaffilmark{10,11}}

\author{Pasquale Mazzotta\altaffilmark{12,13}}

\author{Alastair J. R. Sanderson\altaffilmark{6}}

\author{Daniel P. Marrone\altaffilmark{14,**}}


\altaffiltext{1}{Argelander-Institut f\"ur Astronomie, Universit\"at Bonn, Auf
  dem H\"ugel 71, 53121 Bonn, Germany}
\altaffiltext{2}{Academia Sinica Institute of Astronomy and Astrophysics, P.O. Box 23-141,
  10617 Taipei, Taiwan}
\altaffiltext{3}{Astronomical Institute, Tohoku University, Aramaki, Aoba-ku, Sendai, 980-8578, Japan}
\altaffiltext{4}{Max-Planck-Institut f\"ur extraterrestrische Physik, Giessenbachstra\ss e, 85748 Garching, Germany}
\altaffiltext{5}{University of Maryland, Baltimore County, 1000 Hilltop
  Circle, Baltimore, MD 21250, USA}
\altaffiltext{6}{School of Physics and Astronomy, University of
               Birmingham, Edgbaston, Birmingham, B152TT, UK}
\altaffiltext{7}{CEA, CEA-Saclay, 91191 Gif-sur-Yvette Cedex, France}
\altaffiltext{8}{SISSA/ISAS, via Beirut 4, 34014 Trieste, Italy}
\altaffiltext{9}{Department of Physics and Astronomy, University of Victoria,
  3800 Finnerty Road, Victoria, BC Canada}
\altaffiltext{10}{Department of Physics and Michigan Center for Theoretical
  Physics, University of Michigan, Ann Arbor, MI 48109, USA}
\altaffiltext{11}{Departments of Physics and Astronomy, University of
  California, Berkeley, CA 94720, USA}
\altaffiltext{12}{Dipartimento di Fisica, Universita di Roma ``Tor Vergata,''
  Via della Ricerca Scientifica 1, I-00133 Rome, Italy}
\altaffiltext{13}{Harvard-Smithsonian Center for Astrophysics, 60 Garden
  Street, Cambridge, MA 02138, USA}
\altaffiltext{14}{Kavli Institute for Cosmological Physics, Department of Astronomy and Astrophysics, University of Chicago, Chicago, IL 60637, USA}
\altaffiltext{**}{Hubble Fellow}


\begin{abstract}
  We compare X-ray hydrostatic and weak-lensing mass estimates for a sample of
  12 clusters that have been observed with both \emph{XMM-Newton} and
  Subaru. At an over-density of $\Delta=500$, we obtain $1-M^{\rm X}/M^{\rm
    WL}=0.01 \pm 0.07$ for the whole sample.  We also divided the sample into
  undisturbed and disturbed sub-samples based on quantitative X-ray
  morphologies using asymmetry and fluctuation parameters, obtaining $1-M^{\rm
    X}/M^{\rm WL}=0.09 \pm 0.06$ and $-0.06 \pm 0.12$ for the undisturbed and
  disturbed clusters, respectively. In addition to non-thermal pressure
  support, there may be a competing effect associated with adiabatic
  compression and/or shock heating which leads to overestimate of X-ray
  hydrostatic masses for disturbed clusters, for example, in the famous merging
  cluster A1914. Despite the modest statistical significance of the mass
  discrepancy, on average, in the undisturbed clusters, we detect a clear
  trend of improving agreement between $M^{\rm X}$ and $M^{\rm WL}$ as a
  function of increasing over-density, $M^{\rm X}/M^{\rm WL}=(0.908 \pm
  0.004)+(0.187 \pm 0.010) \cdot \log_{10} (\Delta/500)$. We also examine the
  gas mass fractions, $f_{\rm gas}=M^{\rm gas}/M^{\rm WL}$, finding that they
  are an increasing function of cluster radius, with no dependence on
  dynamical state, in agreement with predictions from numerical simulations.
  Overall, our results demonstrate that \emph{XMM-Newton} and Subaru are a
  powerful combination for calibrating systematic uncertainties in cluster
  mass measurements.
\end{abstract}


\keywords{cosmology: observations --- galaxies: clusters: general ---
  galaxies: clusters: individual (Abell 1914) --- gravitational lensing: weak
  --- surveys --- X-rays: galaxies: clusters}



\section{Introduction}

The mass function of galaxy clusters depends on both the matter density and
the expansion history of the universe.  Indeed, clusters provided early
evidence for a low density universe (White et al.  1993). Clusters, as
powerful tools to constrain cosmological parameters (e.g., Zhang \& Wu 2003;
Balogh et al. 2006; Henry et al. 2009), currently receive much attention as a
potential probe of the dark energy equation of state parameter ($w=p/\rho$,
where $\rho$ is the energy density and $p$ is the pressure), through the
evolution of the mass function. (e.g., Vikhlinin et al. 2009a, 2009b). In
addition, gas mass fraction measurements also potentially provide an important
cosmological probe, under the assumption that gas mass fractions do not evolve
with redshift (e.g., Vikhlinin et al.  2002; Allen et al. 2004, 2008; Mantz et
al. 2010). Upcoming galaxy cluster surveys will shortly deliver huge amounts
of multi-wavelength data, e.g., from Subaru/Hyper-Suprime-Cam, \emph{eROSITA},
\emph{PLANCK}, and South Pole Telescope (SPT). To achieve good control over
systematic errors in cosmological measurements, for example of $w$, based on
these surveys, it is crucial to understand cluster mass estimates.

The density and temperature distributions of the hot X-ray emitting gas within
galaxy clusters can be used to estimate the total mass of the cluster. Since
the acceleration of the gas is still not well understood, it is assumed that
the acceleration terms are negligible when estimating cluster masses from
X-ray data; the resulting mass estimates thus invoke the assumption of
hydrostatic equilibrium (H.E.), and are hereafter referred to as ``X-ray
hydrostatic mass estimates''. Such mass estimates also assume that the total
pressure is dominated by the thermal pressure of the gas.

Numerical simulations (e.g., Evrard 1990; Lewis et al. 2000; Rasia et al.
2006; Nagai et al.  2007; Piffaretti \& Valdarnini 2008, PV08, hereafter;
Jeltema et al.  2008; Lau et al.  2009) have pointed out that X-ray
hydrostatic mass estimates may underestimate cluster mass, and that this
effect is most pronounced for clusters classified as disturbed, based on their
X-ray morphology. Therefore, measurements of cosmological parameters based on
the X-ray-measured cluster mass function or the X-ray-measured gas mass
fraction may be biased. It is thus crucial, as a minimum, to calibrate
observationally both the X-ray-measured cluster masses and gas mass fractions
for a representative cluster sample. This opportunity is offered by measuring
cluster masses using both X-ray and gravitational lensing data.

X-ray and gravitational lensing mass measurements have complementary
advantages and disadvantages. The $n_e^2$ dependence of the X-ray emissivity
of the intracluster medium (ICM) helps guard against projection effects due to
mass along the line of sight through the cluster. However, as discussed above,
X-ray analysis requires assumptions to relate electromagnetic radiation to a
model of the mass distribution. An ideal case is to use an approach that is
insensitive to the dynamical state to estimate the cluster mass.
Gravitational lensing fulfills this requirement because the lensing signal is
insensitive to the physical state and nature of the deflecting matter
distribution. Lensing is, however, prone to projection effects because it is
sensitive to all mass along the line of sight through the cluster (e.g.,
Hoekstra 2001). It is thus of paramount importance to combine these two
techniques to develop a thorough understanding of cluster mass measurements.
Since the 1990's, the use of X-ray and lensing observations has been proposed to
test deviations from H.E. and extra pressure support (e.g., Miralda-Escude \&
Babul 1995; Wu \& Fang 1996; Squires et al 1996; Allen 1998; Zhang et
al. 2005, 2008; Mahdavi et al. 2008; Richard et al. 2010). Most earlier
studies employed "target of opportunity" mode --- i.e., considered clusters
with available data --- making the studies biased by design.

The Local Cluster Substructure Survey (LoCuSS\footnote{\sf
  http://www.sr.bham.ac.uk/locuss}; G. P. Smith et al. in preparation; Zhang
et al. 2008; Haines et al. 2009a, 2009b; Sanderson et al. 2009; Marrone et al.
2009; Okabe et al. 2010a; Richard et al. 2010) is a systematic
multi-wavelength survey of X-ray luminous ($L_{\rm X,\,0.1-2.4keV}{\ge}2 \cdot
10^{44}{\rm erg~s}^{-1}$) galaxy clusters at $0.15\simlt z \simlt 0.3$
selected from the \emph{ROSAT} All-Sky Survey (RASS; Ebeling et al. 2000;
B\"ohringer et al. 2004) in a manner blind to the dynamical status. As a first
step toward a comprehensive X-ray/lensing study we assembled a sample of 19
clusters with archival \emph{XMM-Newton} observations, and weak-lensing mass
measurements in the literature (Zhang et al. 2008). The mean weak-lensing mass
to X-ray hydrostatic mass ratio was found to be $\langle M^{\rm WL}/M^{\rm
  X}\rangle=1.09\pm0.08$ at an over-density of $\Delta=500$ with respect to
the critical density. Mahdavi et al. (2008) found the average X-ray to
weak-lensing mass ratio to be $0.78 \pm 0.09$ at $\Delta=500$, being
consistent with the results in Zhang et al. (2008). The uncertainties in the
above analysis was dominated by measurement errors, in particular, the lensing
masses drawn from the literature based on early lensing data (Bardeau et
al. 2007; Dahle 2006), with typical seeing of ${\rm FWHM}{\gs}0.^{\prime
  \prime}8$, using four different cameras on three different ${\le}3.6$-m
telescopes, with fields of view (FOVs) spanning $r{\sim}0.7{-}2\,{\rm Mpc}$ at
$z \sim 0.2$. The next step in our program is to employ uniform high-quality
weak-lensing data from our own dedicated observing program with
Subaru/Suprime-CAM. Recently, the first batch of weak-lensing mass
measurements have become available for 22 clusters based on data taken in good
conditions (${\rm FWHM}{\simeq}0.^{\prime \prime}6$) through two-filters with
details outlined in Section~\ref{s:lens} (Okabe et al. 2010a; Okabe \& Umetsu
2008). These clusters were selected based on observability from Mauna Kea on
nights allocated to LoCuSS, and are thus unbiased with respect to their X-ray
properties.  Archival \emph{XMM-Newton} data are available for 12 of the 22
clusters --- see Zhang et al. (2008) for details.  In this work, we therefore
compare X-ray (based on the assumption of H.E.)  and weak-lensing mass
estimates for these 12 clusters, and also compare our observational results
with predictions from numerical simulations. It is worth noting that
Subaru/Suprime-CAM provides an excellent match to the \emph{XMM-Newton} FOV,
and also that these Subaru weak-lensing mass estimates have optimal precision
in the density contrast range of $500\le \Delta \le 2000$ (Okabe et
al. 2010a), again, well matched to the \emph{XMM-Newton} X-ray estimates.

The outline of this paper is as follows.  In Section~\ref{s:data} we briefly
describe the weak-lensing and X-ray analysis. The results are presented in
Section~\ref{s:bias}, and discussed in Section~\ref{s:dis}. We summarize our
conclusions in Section~\ref{s:con}. Throughout the paper, we assume
$\Omega_{\rm m}=0.3$, $\Omega_\Lambda=0.7$, and
$H_0=70$~km\,s$^{-1}$\,Mpc$^{-1}$.  Confidence intervals correspond to the
68\% confidence level. Unless explicitly stated otherwise, we apply the
Orthogonal Distance Regression package
(ODRPACK~2.01\footnote{http://www.netlib.org/odrpack and references therein},
e.g., Boggs et al. 1987) taking into account measurement errors on both
variables to determine the parameters and their errors of the fitting.

\section{Data analysis} 
\label{s:data}

The 12 clusters in our sample are listed in Table~\ref{t:cen}.

\subsection{weak-lensing analysis}
\label{s:lens}

The Subaru $i^{\prime}$- and $V$-band imaging data and detailed weak-lensing
analysis are described in Section~2 in Okabe et al. (2010a), in which the
$i^{\prime}$-band data are used for shape measurements, and the $V$-band data
are used to remove foreground and cluster galaxies. Excluding unlensed
galaxies from the background galaxy catalog is of prime importance for
accurate weak-lensing mass estimates, especially at higher over-densities.
This so called dilution bias increases as a function of the density contrast
$\Delta$, and can cause $M^{\rm WL}_{500}$ and $M^{\rm WL}_{2500}$ to be
biased low by $\sim20\%-50\%$ (Okabe et al. 2010a). Our two-filter lensing data
allow us to construct secure {\it red+blue} background galaxy samples, defined
as faint galaxies with colors that are redder and bluer than the cluster
red-sequence by a minimum color offset.  This strategy typically reduces the
dilution bias to per cent level (Okabe et al. 2010a). The mean redshift for
the background galaxy catalog is estimated by matching the magnitudes and
colors of background galaxies to the COSMOS photometric redshift catalog
(Ilbert et al. 2009).  More precisely, the mean redshift is computed as a lens
weighted average over the redshift distribution, $dP_{\rm WL}/dz$, which is
defined by $\langle D_{\rm ls}/D_{\rm s}\rangle=\int_{z_{\rm d}}dz d P_{\rm
  WL}/dz D_{\rm ls}/D_{\rm s}$, where $D_{\rm s}$ and $D_{\rm ls}$ are the
angular diameter distances to source and between lens and source,
respectively.

Galaxy cluster mass distributions are often modeled as Navarro, Frenk, and
White (NFW; Navarro et al.  1997) halos or singular isothermal spheres
(SIS). In weak-lensing studies, the distortion signal is used to constrain the
model parameters. Okabe et al. (2010a) have shown that the NFW model fits the
lensing distortion profiles well in both statistical studies and their
analysis of individual clusters. The SIS model is statistically inadequate to
describe stacked tangential distortion profiles with pronounced radial
curvatures, and is rejected at $6 \sigma$ and $11\sigma$ level, respectively,
for their two sub-samples with the virial masses in the ranges of $< 6 \times
10^{14} h^{-1} M_{\odot}$ and $\ge 6 \times 10^{14} h^{-1} M_{\odot}$. It is
also important to note that the mean ratio of masses obtained from SIS and NFW
models is $0.70\pm0.05$ at $r^{\rm WL}_{500}$ for our sample of 12 clusters.

The projected mass distribution can also be obtained using a model-independent
approach, the so-called $\zeta_{\rm c}$-statistics method (Fahlman et
al. 1994; Clowe et al. 2000), which is complementary to the tangential shear
fit method. The $\zeta_{\rm c}$-statistics method measures the discrete
integration of averaged tangential distortions of source galaxies outside
given radii. The $\zeta_{\rm c}$-statistics method is thus less sensitive to
the detailed structure of clusters on small scales than using models fitted to
the full tangential shear data. Okabe et al.'s $\zeta_{\rm c}$-based
model-independent projected masses are in good agreement with the projected
masses computed from the NFW-based models at $\Delta=500$, but not with the
SIS-based models.

We also compared the NFW-based spherical mass estimates for our sample of 12
clusters with the spherical mass estimates based on deprojecting the
$\zeta_{\rm c}$-based model-independent masses. In the latter case, the
spherical mass was obtained by assuming an NFW profile. The spherical masses
using the $\zeta_{\rm c}$-statistics and NFW models are statistically
consistent, given their mean ratios of $1.00\pm0.08$, $0.96\pm0.07$, and
$1.02\pm0.07$, at $r^{\rm WL}_{500}$, $r^{\rm WL}_{1000}$, and $r^{\rm
  WL}_{2500}$, respectively, weighted by the inverse square of the errors. For
two clusters, A383 and A2390, the masses from the NFW models are lower than
the deprojected masses from the $\zeta_{\rm c}$-statistics method due to
substructures in the cluster cores.

Based on the above tests, our joint analysis uses the NFW masses obtained from
the tangential distortion profiles in Okabe et al. (2010a). The
three-dimensional spherical cluster masses (see Table~\ref{t:mass}), $M^{\rm
  WL}_\Delta$, within a sphere of radius $r_\Delta$, is derived following the
NFW model, $\rho \propto r^{-1}(1+c_\Delta r/r_{\Delta})^{-2}$, where
$c_\Delta$ is the concentration parameter.

The weak-lensing analysis includes both statistical error and errors of the
photometric redshifts of source galaxies in mass measurements. The latter is
estimated by bootstrap re-sampling to match the COSMOS catalog.
Leauthaud et al. (2010) found that the latter mainly matters for galaxies
close to the lens. The typical $1\sigma$ total uncertainty on the weak-lensing
mass estimates is $\sim 12\%-21\%$ at $\Delta=500$ with four exceptions, i.e.,
RXJ2129.6+0005, 39\%; A68, 41\%; A115 (south), 53\%; and Z7160, 42\%.

Hoekstra (2001) pointed out that the uncertainty in weak-lensing mass
estimates of clusters, caused by distant large-scale structures (uncorrelated)
along the line of sight is fairly small for deep observations ($20<R<26$) of
massive clusters at intermediate redshifts. The typical $1\sigma$ relative
uncertainty is about 6\% if the lensing signal is measured out to $1.5
h^{-1}_{50}$ Mpc. All 12 clusters in our sample are massive clusters
($>5$~keV) at intermediate redshifts ($0.15\le z \le0.3$) with deep Subaru
observations to $\sim 26$ mag. The Subaru FOV covers the entire cluster up to
a few Mpc for our sample. Therefore, the mean mass uncertainty caused by
large-scale structures due to projection should be no more than 6\%. Therefore
we neglect this error in the mass estimates.

\subsection{X-ray analysis}
\label{s:xray}

The X-ray analysis was carried out independently from the weak-lensing
analysis. The description of \emph{XMM-Newton} data and X-ray mass modeling is
given in Section~2 in Zhang et al. (2008), and more details on the data
reduction are given in Section~2 in Zhang et al. (2006, 2007). The radial
temperature profile was measured using spectral data including
deprojection. The X-ray spectrum measuring the global temperature was used to
calculate the cooling function for the conversion from X-ray surface
brightness profile to electron number density profile, in which we included
both deprojection and \emph{XMM-Newton} point-spread-function corrections. The
gas mass $M^{\rm gas} (\le r^{\rm WL}_{\Delta})$ was derived by integrating
the electron number density, which was fitted by a double-$\beta$ model, and
assuming $\mu_{\rm e}=1.17$. The X-ray hydrostatic mass $M^{\rm X}(\le r^{\rm
  WL}_{\Delta})$ was measured from the temperature profile and electron number
density profile assuming spherical symmetry and H.E.. Unless explicitly stated
otherwise, both the X-ray gas mass and the hydrostatic mass are measured
within the radius $r_\Delta^{\rm WL}$ obtained in the weak-lensing
analysis. The typical $1\sigma$ uncertainty on the X-ray hydrostatic mass is
$\sim 18\%-30\%$ at $\Delta=500$ with no exceptions (see Table~\ref{t:mass}).
     
\subsection{Selection of cluster centers}

The cluster centers were derived independently in the weak-lensing and X-ray
analysis.  In the weak-lensing analysis, we follow Okabe et al. (2010a, see
their Section~3.2) and adopt the position of the brightest cluster galaxy
(BCG) as the cluster center. This is motivated by the coincidence of the peak
of the two-dimensional cluster mass distribution with the BCG position, and
the strong gravitational lensed images centered close to the BCG position in
some of the clusters (Richard et al. 2010).

In the X-ray analysis in Zhang et al.  (2008), the cluster center is
determined using the flat fielded X-ray image in the 0.7-2~keV band as
follows. The procedure is initiated by deriving the first flux-weighted center
within a 1$^{\prime}$ aperture centered at the peak of the cluster X-ray
emission. Iteratively, we re-derive the flux-weighted center but within the
aperture, which is 1$^{\prime}$ larger than the previous one and centered at
the previous flux-weighted center, till the coordinates of the flux-weighted
center do not vary anymore. The iteration is less than 10 times to fulfill
the goal. The final flux-weighted center is taken as the X-ray cluster center.

We list the lensing and X-ray centers in Table~\ref{t:cen}; they agree to
within $0.14 r_{2500}$ for all except two clusters, namely A1914 and
RXCJ2337.6+0016. We therefore test the sensitivity of the lensing and X-ray
analysis to the choice of center, finding that the systematic error is small.
For example, if we adopt the BCG as the center of the X-ray analysis of A1914
then the hydrostatic mass estimates change by 1\%, 10\%, and 3\% at $\Delta
=$500, 1000, and 2500, respectively. Similarly, if we adopt the X-ray center
as the center of the lensing analysis of the same cluster, then the lensing
mass estimates change by 0.1\%, 0.2\%, and 0.4\% also at $\Delta = $500, 1000,
and 2500, respectively. The X-ray analysis of the two clusters was therefore
revised using the weak-lensing determined cluster center.

\subsection{X-ray morphology}

The X-ray morphology of each cluster was determined by Okabe et al. (2010b) by
calculating asymmetry ($A$) and fluctuation parameters ($F$; Conselice 2003)
from the \emph{XMM-Newton} X-ray images. In summary, the asymmetry parameter
is defined as $A=(\sum_{ij} |I_{ij}-R_{ij}|)/\sum_{ij} I_{ij}$, the normalized
sum of the absolute value of the flux residuals.  $I_{ij}$ is the element of
the \emph{XMM-Newton} MOS1+MOS2 image in the 0.7-2.0~keV band, which is flat
fielded, point source subtracted and re-filled assuming a Poisson
distribution, and binned by $4^{\prime \prime} \times 4^{\prime
  \prime}$. $R_{ij}$ is the element of the image derived by rotating the above
image by $180^\circ$. We take into account the position resolution of
\emph{XMM-Newton} by allowing the cluster center falling into any neighboring
pixel of the $r\le 4^{\prime \prime}$ circle centered at the BCG and include
this error in quadrature in calculating $A$. Dynamically immature clusters
often show both an asymmetric X-ray morphology and an offset between
weak-lensing and X-ray centers. Therefore $A$ is very sensitive to cluster
dynamical state. The fluctuation parameter is given by $F=(\sum_{ij}
I_{ij}-B_{ij})/\sum_{ij} I_{ij}$, in which $B_{ij}$ is the element in the
smoothed image. This parameter describes the degree of deviations from the
smoothed distribution. We measure the errors of $A$ and $F$ assuming a Poisson
noise computed within a radius of $r^{\rm WL}_{500}$, excluding CCD gaps and
bad pixels.

The $F$ versus $A$ plane is divided into four quadrants with cuts at $A=1.1$
and $F=0.05$ in Figure~1 in Okabe et al. (2010b). Our sample of 12 clusters
occupy these quadrants as follows: (1) RXCJ2129, A209, A383, A1835, and A2390
have both low $A$ and low $F$; (2) A2261 and A1914 have high asymmetry
parameters; (3) A68, RXCJ2337, A267, and Z7160 have high fluctuation
parameters; and (4) A115 (south) has both high $A$ and high $F$. The clusters
falling into quadrant (1) are defined as dynamically undisturbed clusters, and
the remaining clusters as disturbed clusters because high $A$ and/or high $F$
indicates that the cluster is still dynamically young. It is worth noting that
the five undisturbed clusters have $\le 0.06 r_{2500}$ offset between the
X-ray cluster center and the weak-lensing cluster center which is the BCG
position.

\section{Results} 
\label{s:bias}

\subsection{X-ray hydrostatic mass versus weak-lensing mass}
\label{s:biasobs}

\begin{figure*}
\begin{center}
\includegraphics[angle=270,width=8cm]{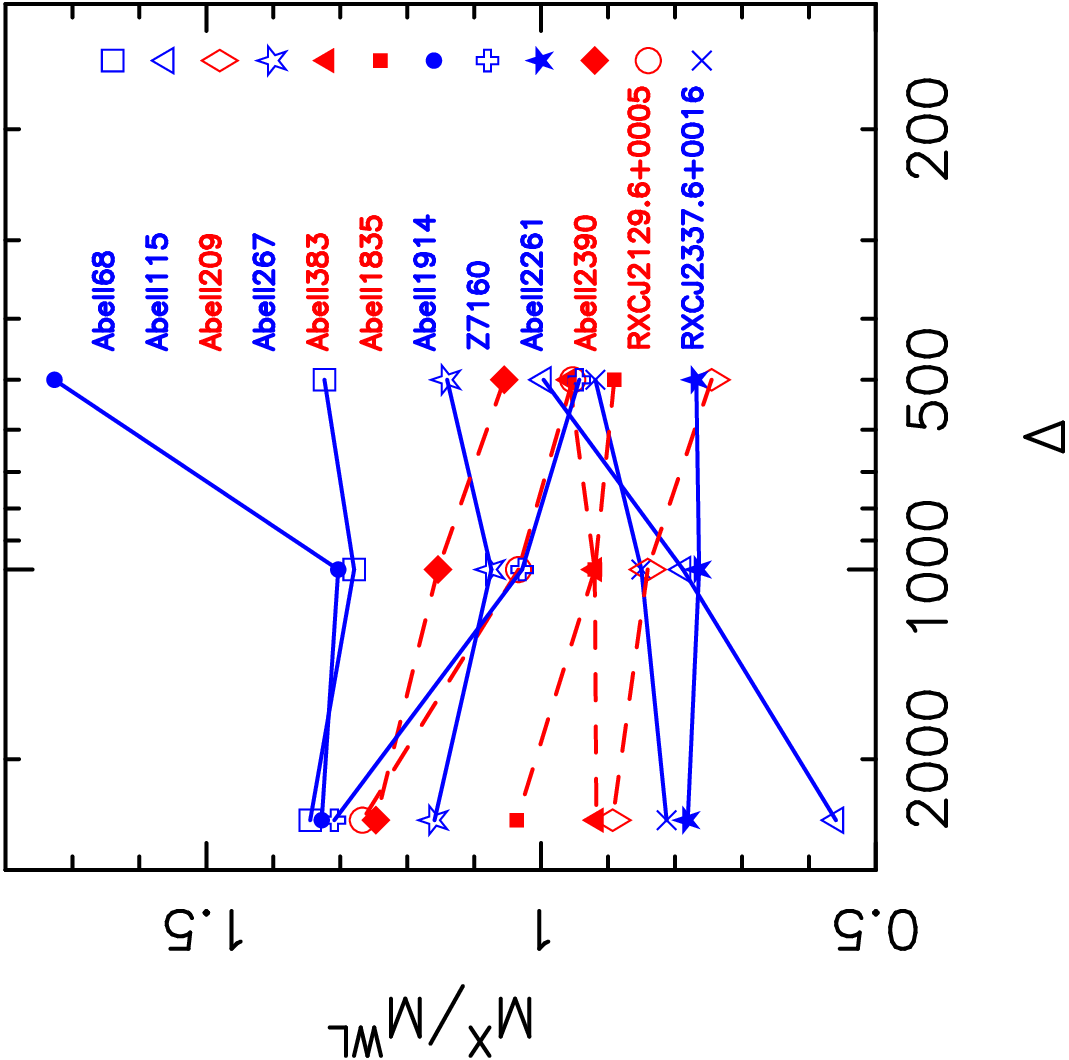}
\includegraphics[angle=270,width=8cm]{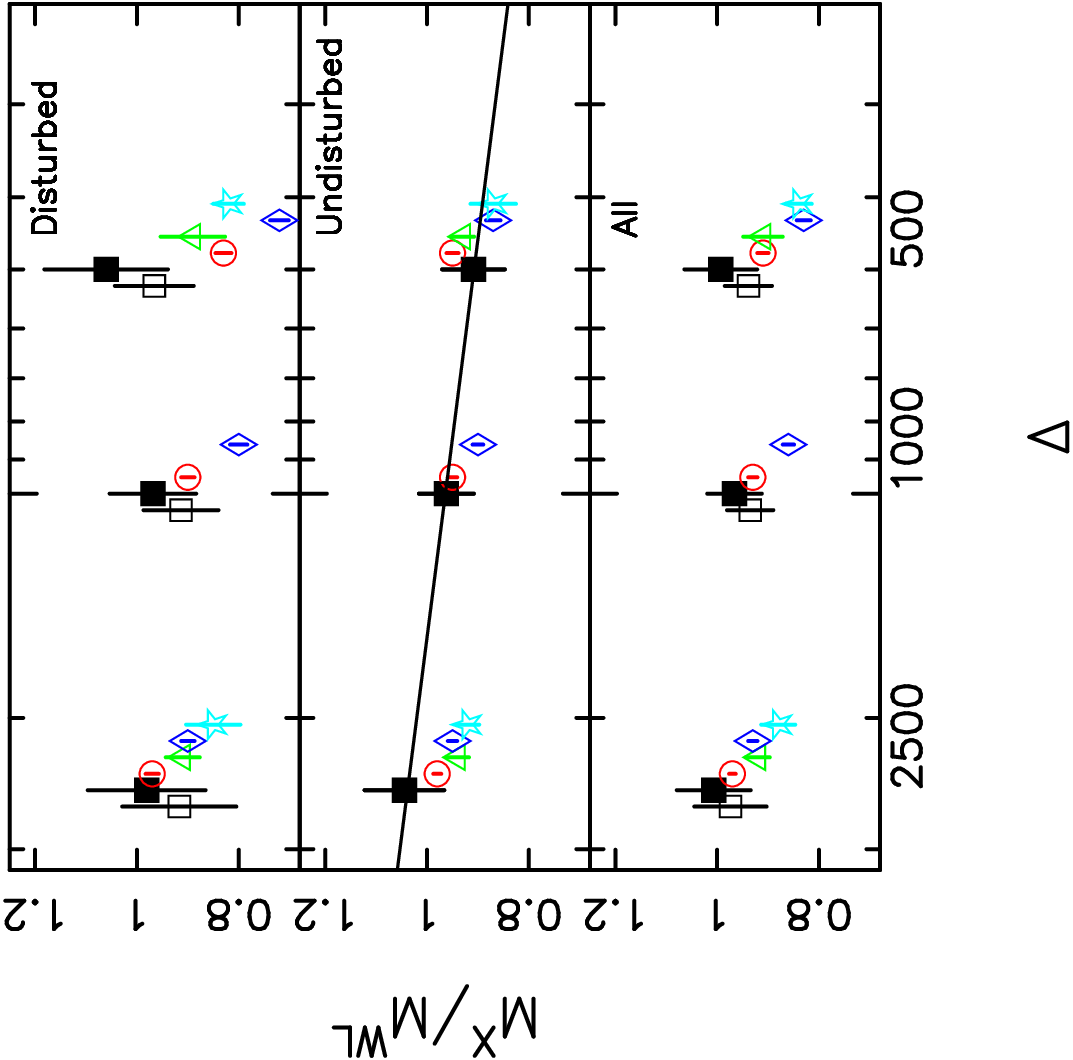}
\end{center}
\caption{{\em Left panel} : X-ray hydrostatic to weak-lensing mass ratios as a
  function of density contrast for individual clusters, in which red dashed
  and blue solid lines denote 
  undisturbed and disturbed clusters, respectively. {\em Right panel}: X-ray
  to weak-lensing mass ratios for all clusters (bottom), undisturbed clusters
  (middle), and disturbed clusters (top) from our sample including (filled
  black boxes) and excluding (open black boxes) A1914, together with the X-ray
  hydrostatic mass to true mass ratios from numerical simulations (PVge2keV,
  red circles using $T_{\rm MW}$, blue diamonds using $T_{\rm SL}$; Lau et
  al. 2009, green triangles using quasi-$T_{\rm MW}$; Nagai et al.  2007,
  light blue stars using $T_{\rm SL}$). The data points at each density
  contrast are off by 0.022~dex for clarity. We also show the best fit of the
  X-ray hydrostatic to weak-lensing mass ratio as a function of density
  contract for the five undisturbed clusters, $M^{\rm X}/M^{\rm WL}=(0.908 \pm
  0.004)+(0.187 \pm 0.010) \cdot \log_{10} (\Delta/500)$. (A color version of
  this figure is available in the online journal.)}
\label{f:mx_to_mwl}
\end{figure*}

\begin{figure*}
\begin{center}
\includegraphics[angle=270,width=5.4cm]{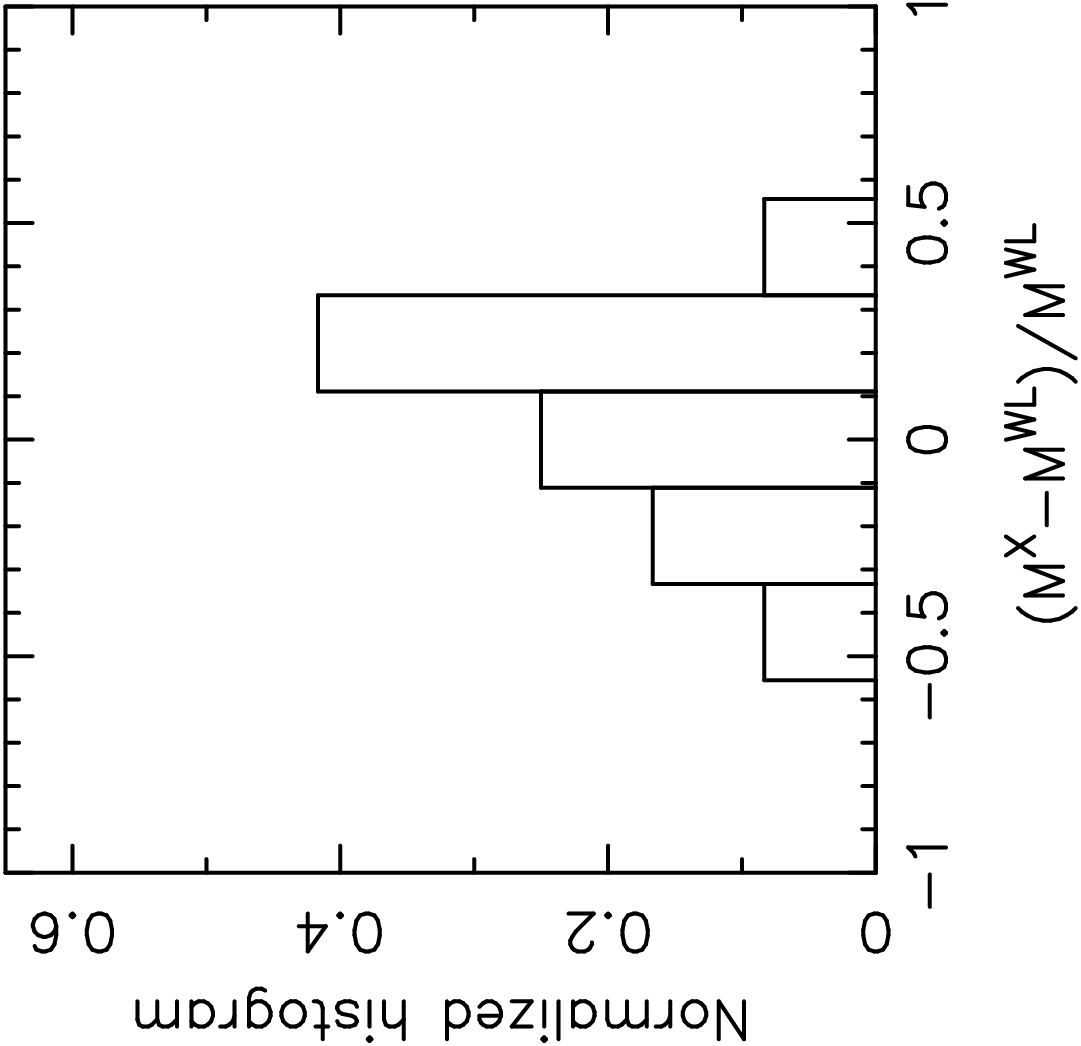}
\includegraphics[angle=270,width=5.4cm]{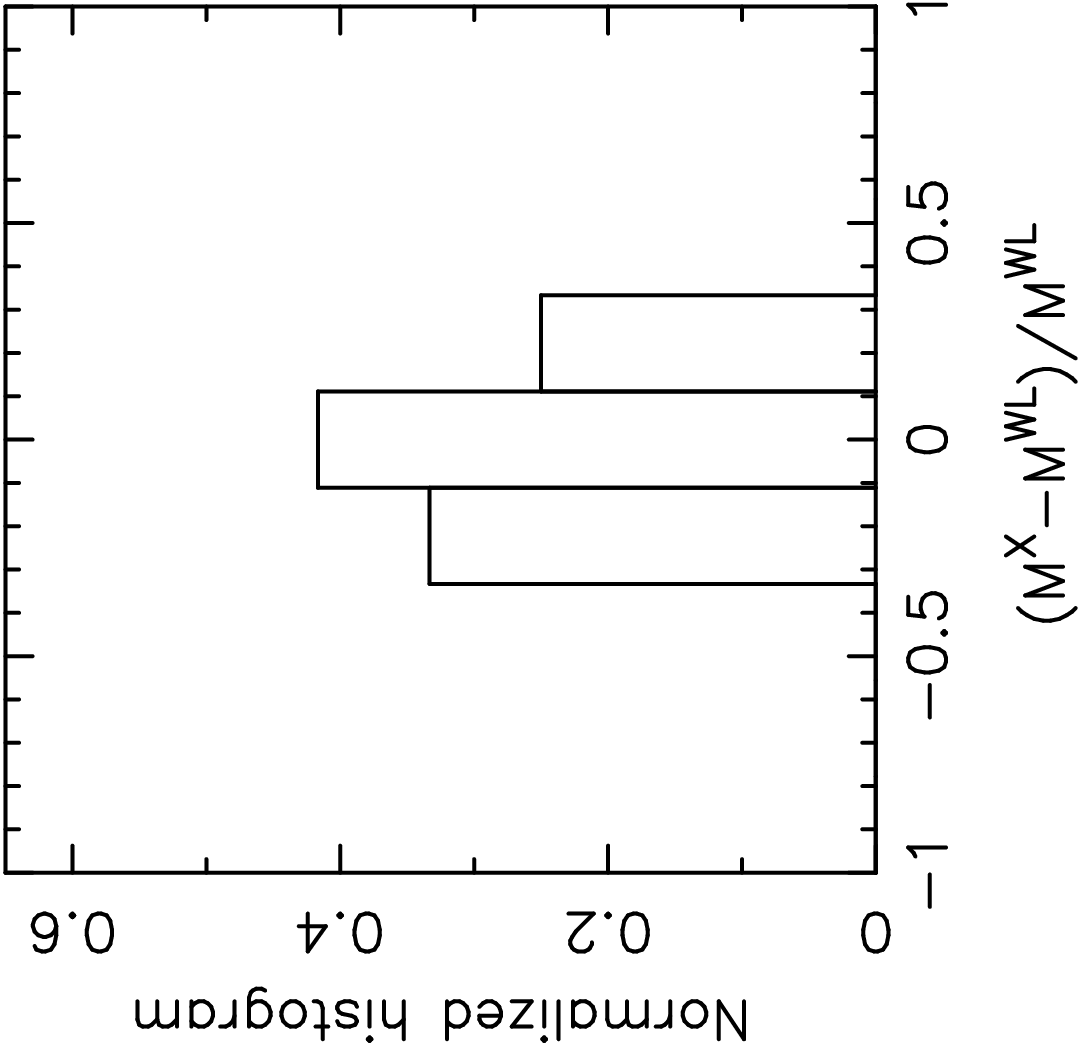}
\includegraphics[angle=270,width=5.4cm]{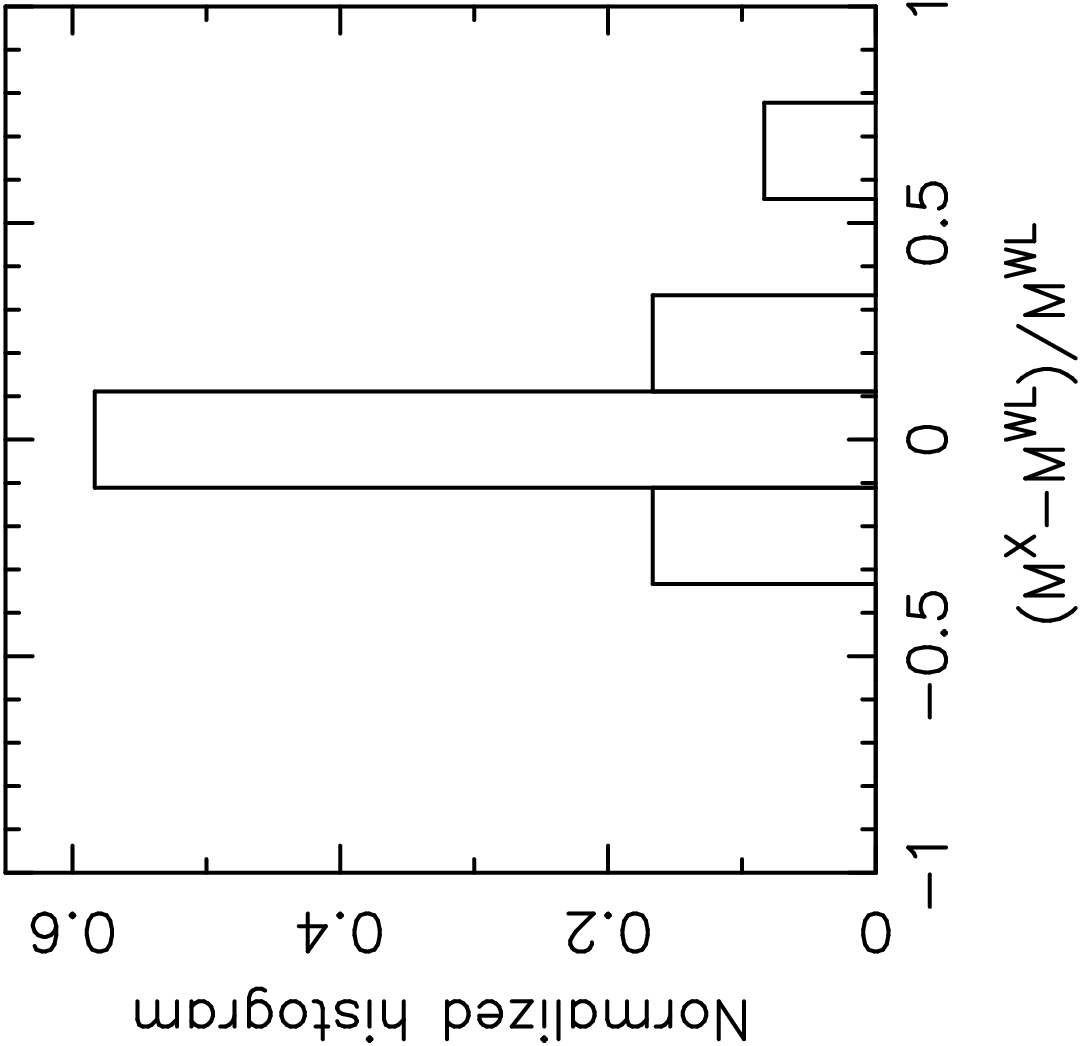}
\end{center}
\caption{Normalized histograms of $(M^{\rm X}-M^{\rm WL})/M^{\rm WL}$ for all
  12 clusters at $\Delta=2500$ (left panel), $\Delta=1000$ (middle panel), and
  $\Delta=500$ (right panel).}
\label{f:hist}
\end{figure*}

The comparison between X-ray and weak-lensing masses for individual clusters
is shown in the left panel of Figure~\ref{f:mx_to_mwl}. The X-ray to
weak-lensing mass ratios vary in the range of $\sim 0.55-1.72$. Undisturbed
clusters have X-ray to weak-lensing mass ratios that generally increase
toward smaller cluster-centric radii (higher over-density).  In contrast,
disturbed clusters show a much greater diversity, with some disturbed clusters
having mass ratios that increase toward larger cluster-centric radius.  The
distribution of $(M^{\rm X}-M^{\rm WL})/M^{\rm WL}$ at $\Delta=500$ (right
panel of Figure~\ref{f:hist}) reveals that the well-known merging cluster
A1914 is a $\sim5\sigma$ outlier based on a naive calculation of the mean mass
ratio for the other 11 clusters.  As pointed out in Section~\ref{s:xray},
A1914 also shows the largest offset ($0.44 r_{2500}$) between the X-ray and
weak-lensing determined cluster centers.

The average X-ray to weak-lensing mass ratios for the full sample and the
undisturbed and disturbed sub-samples were calculated taking into account the
errors in the X-ray and weak-lensing mass estimates --- see
Table~\ref{t:mratio} and the right panel of Figure~\ref{f:mx_to_mwl}. $M^{\rm
  X}/M^{\rm WL}$ is consistent with unity across the full radial range probed
by the data for the full sample of 12 clusters. This also holds for the seven
disturbed clusters, albeit with uncertainties $\sim2\times$ those for the full
sample (as seen in simulations; e.g., Nagai et al. 2007). In contrast,
undisturbed clusters show a gentle decline in $M^{\rm X}/M^{\rm WL}$ to larger
cluster-centric radii --- at $\Delta=500$, the five undisturbed clusters show an
average X-ray hydrostatic mass $9\%\pm6\%$ lower than the average weak-lensing
mass.

Interestingly, given the apparent trend of $M^{\rm X}/M^{\rm WL}$ with radius,
$M^{\rm X}/M^{\rm WL}$ is consistent with unity for the full sample at
$\Delta=500$.  As noted above, A1914 --- the most extreme of the disturbed
clusters has a mass ratio of $\sim1.7$ at $\Delta=500$ (see Table~\ref{t:mass}
and Figure~\ref{f:mx_to_mwl}).  We therefore investigate the extent to which
A1914 might be dominating the results.  We recalculated $M^{\rm X}/M^{\rm WL}$
for the full sample and the disturbed clusters excluding A1914 (right panel of
Figure~\ref{f:mx_to_mwl} and Table~\ref{t:mratio}), finding that the average
X-ray hydrostatic mass is now lower than the average weak-lensing mass just by
$6\%\pm5\%$.  Within the uncertainties, our result that X-ray hydrostatic and
weak-lensing masses agree for the full sample across the full radial range
probed, is therefore insensitive to the inclusion/exclusion of A1914.
However, there is evidence for shock heating of the ICM in the entropy map of
A1914. In addition to non-thermal pressure support (likely the main reason for
X-ray hydrostatic masses being underestimated for undisturbed clusters), there
may also be a competing effect associated with adiabatic compression and/or
shock heating of the intracluster gas which leads to an overestimate of X-ray
hydrostatic masses for some disturbed clusters. We therefore argue that robust
constraints on the bias of X-ray hydrostatic mass estimates for precision
cluster cosmology requires {\it both statistically large and complete
  (unbiased)} samples in order to sample the full range of physical processes
at play within the underlying cluster population.

We also calculated the cumulative probability distribution function of the
X-ray to weak-lensing mass ratio assuming each data point to be a Gaussian
distributed variable and taking into account the errors with 500 Monte Carlo
simulations. The mean and its standard error are listed in
Table~\ref{t:pdf}. Again, a clear trend is found that the average hydrostatic
to weak-lensing mass ratio declines with cluster-centric radius for
undisturbed clusters.  The mass ratios for the full sample and disturbed
clusters are consistent with unity across the full radial range.

To further test our results on undisturbed and disturbed clusters based on our
small sample, we applied the jackknife method to recalculate the average X-ray
to weak-lensing mass ratios at $\Delta=500$.  We randomly removed one system
from the five undisturbed clusters and obtained average X-ray to weak-lensing
mass ratios in the range of 0.875-0.962. Therefore, our finding that the X-ray
hydrostatic mass is on average lower than the weak-lensing mass for
undisturbed clusters is robust. We applied the same procedure to the disturbed
sub-sample and found that the average X-ray to weak-lensing mass ratios are in
the range of 0.966-1.145.  Given the large scatter and measurement errors, it
is therefore unclear whether the average X-ray hydrostatic mass is lower or
higher than the average weak-lensing mass for disturbed systems.

Despite the modest statistical significance, the analysis described above all
points toward a trend of improving agreement between $M^{\rm X}$ and $M^{\rm
  WL}$ as a function of increasing over-density. We therefore proceed a simple
fit to the data, and obtain the following relation: $M^{\rm X}/M^{\rm
  WL}=(0.908\pm0.004)+(0.187\pm0.010)\cdot\log_{10}(\Delta/500)$. These
results are consistent with those of Mahdavi et al.\ (2008), who found an
X-ray to weak-lensing mass ratios of 1.06, 0.96, and 0.85 at $\Delta=2500$,
1000, and 500 respectively, with a typical error bar of 10\%.  Our trend is
slightly shallower than Mahdavi et al.'s result, but is in agreement within
the uncertainties.

Finally, we investigate the issue of scatter.  The undisturbed clusters
present a factor of $\sim2$ less scatter around their average X-ray to
weak-lensing mass ratio than disturbed clusters (Table~\ref{t:mratio}).  A key
question is whether this lower scatter implies lower intrinsic variance within
the undisturbed sub-sample.  We therefore calculated the real variance
following Appendix A of Sanderson \& Ponman (2010) and found that the real
variance for disturbed clusters is $\sim5 - 10\times$ larger than that for
undisturbed clusters at $\Delta=2500$, 1000, and 500 (Table~\ref{t:pdf}).
This confirms that the smaller scatter measured for undisturbed clusters
reflects low intrinsic variance in this cluster population. This result is in
agreement with studies based on numerical simulations, and is fully expected
in both simulations and observations because of the non-smoothness of the gas
distribution, complex thermal- and non-thermal-structure, deviations from
spherical symmetry, etc., that are typical of dynamically immature clusters
(e.g., Poole et al. 2006; Fabian et al. 2008; Zhang et al. 2009).  We also
stress that the fact that disturbed clusters are also typically less spherical
than undisturbed clusters will also render the deprojection of the lensing
signal via fitting a three-dimensional NFW model less valid in disturbed
clusters than in undisturbed clusters.  A thorough investigation of these
issues requires a large complete sample of clusters with deep X-ray and
lensing data.

\subsection{X-ray to weak-lensing mass ratio versus morphology indicators}

We now investigate the dependence of the X-ray to weak-lensing mass ratio on
the morphological parameters, $A$ (asymmetry) and $F$ (fluctuation), and the
concentration of the best-fit NFW halos from Okabe et al. (2010a, 2010b).  As
the mass estimates for individual clusters have large uncertainties, a
detailed quantitative study of the relationship between X-ray to weak-lensing
mass ratio and $A$ and $F$ is beyond the scope of that possible with the
current sample.  We therefore restrict our attention in this section to
general trends that will be worth following up with future larger samples.

\begin{figure*}
\begin{center}
\includegraphics[angle=270,width=5.4cm]{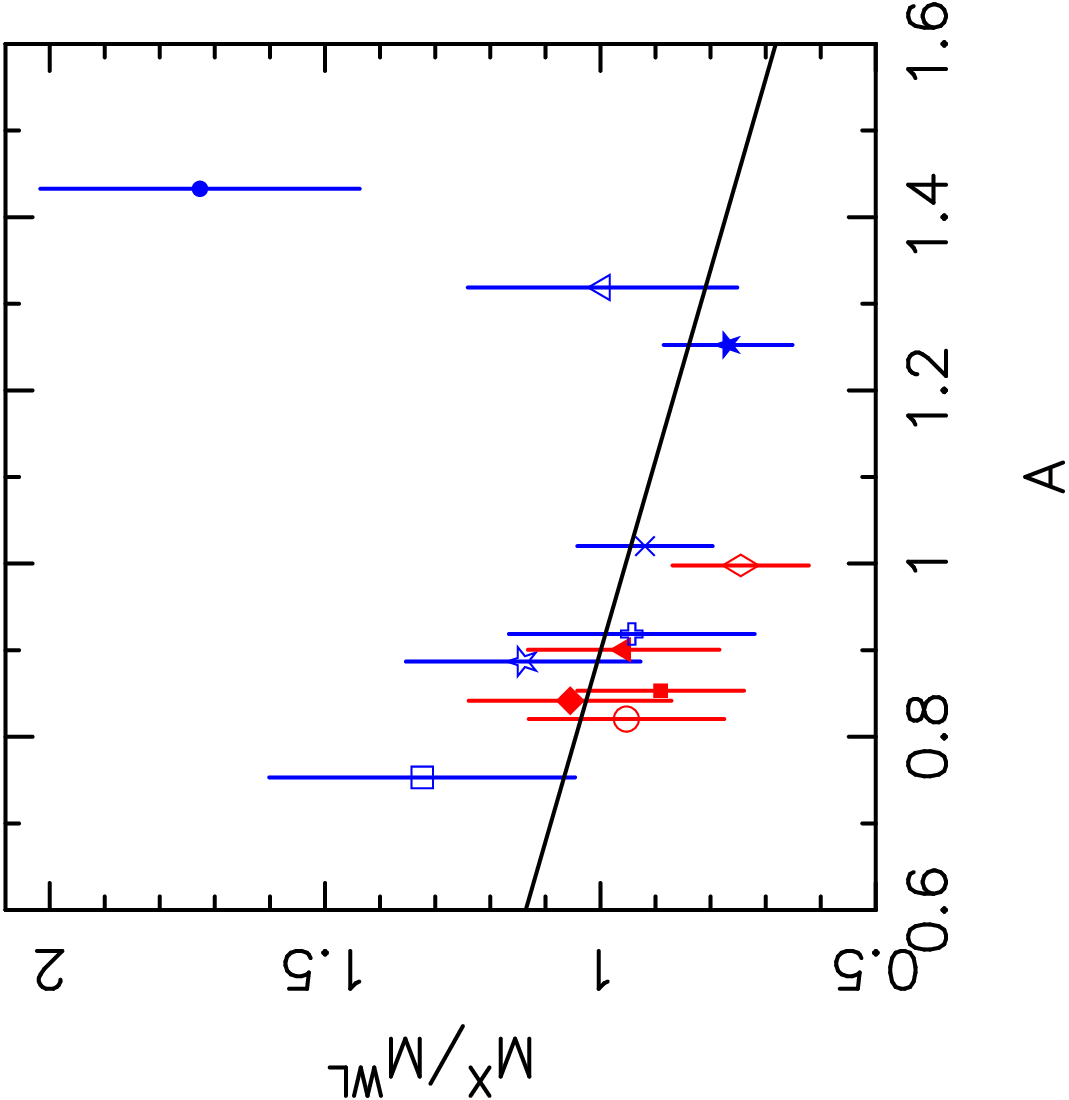}
\includegraphics[angle=270,width=5.4cm]{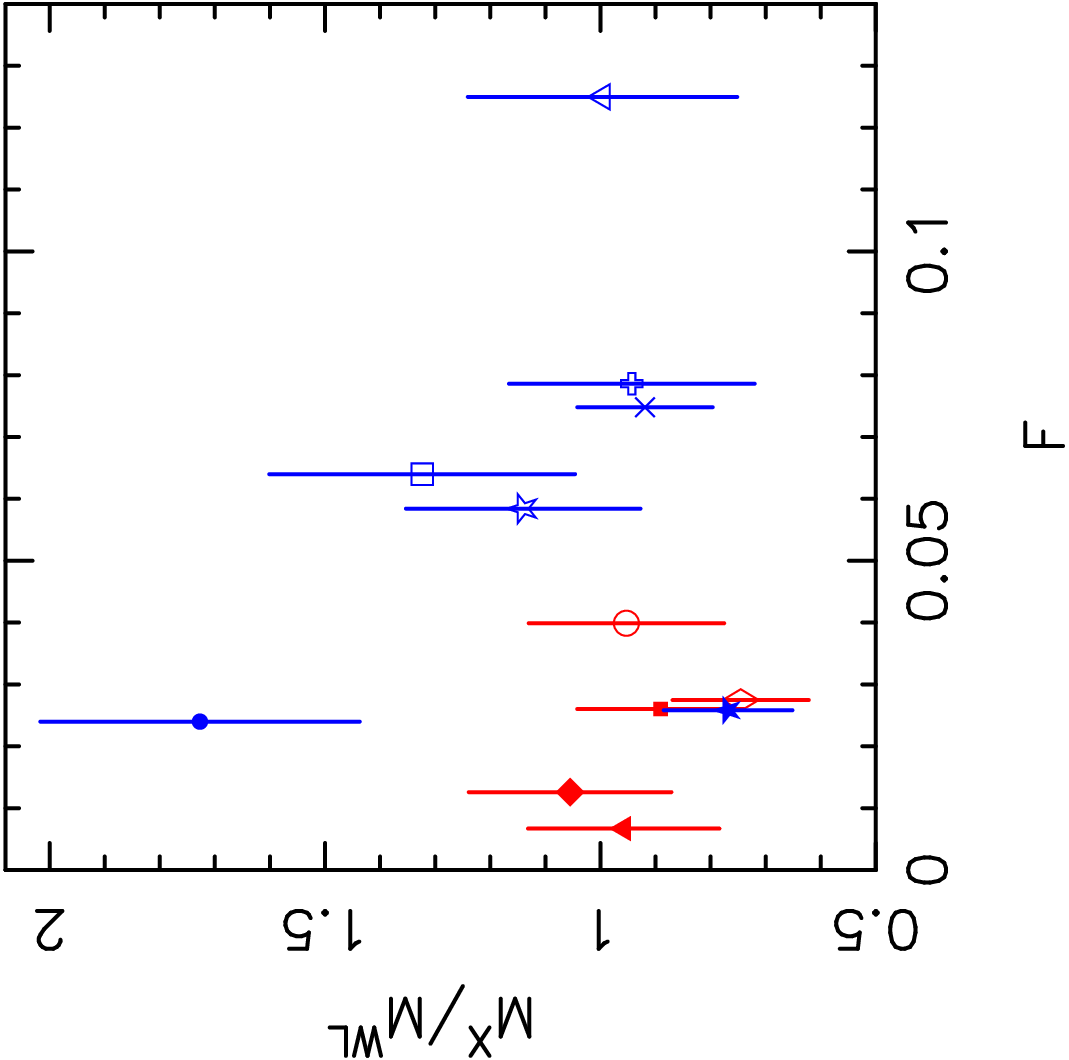}
\includegraphics[angle=270,width=5.4cm]{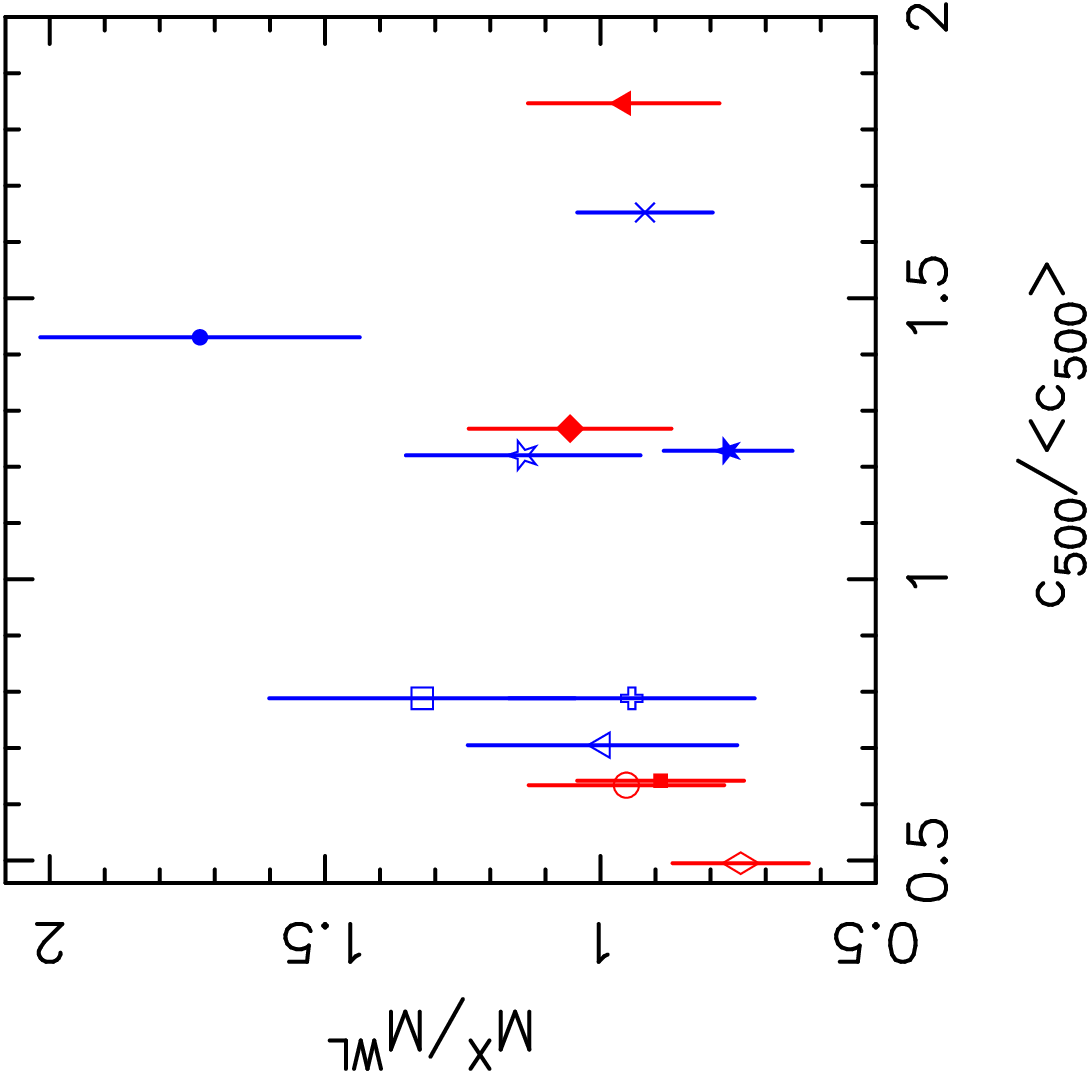}
\end{center}
\caption{X-ray hydrostatic to weak-lensing mass ratio at $\Delta=500$ vs.
  asymmetry parameter (left panel), fluctuation parameter (middle panel), and
  $c_{500}/\langle c_{500} \rangle$ (right panel), respectively. The colors
  and symbols are the same as shown in the left panel of
  Figure~\ref{f:mx_to_mwl}. The best fit with equal weighting of all data
  points except for A1914 (the top rightmost point) is shown in the left
  panel. (A color version of this figure is available in the online journal.)}
\label{f:mbias}
\end{figure*}

First, we plot the X-ray to weak-lensing mass ratio versus the asymmetry and
fluctuation parameters in Figure~\ref{f:mbias}. There is a general trend of
decreasing $M^{\rm X}/M^{\rm WL}$ with increasing asymmetry in the sense that
the most asymmetric clusters have the largest mass discrepancies with $M^{\rm
  X}<M^{\rm WL}$.  A1914 is an obvious outlier from this apparent
anti-correlation.  We therefore exclude A1914 and ignore the observational
error bars in order to fit a ``toy-model'' to the data, obtaining $M^{\rm
  X}/M^{\rm WL}=(1.408\pm0.257)-(0.454\pm0.263)\cdot A$.  We also note that at
fixed asymmetry undisturbed clusters generally have lower $M^{\rm X}/M^{\rm
  WL}$ than disturbed clusters, with undisturbed systems possibly tracing a
steeper and tighter trend than the latter.  However, we caution that these
results are preliminary --- for example, the straight-line fit discussed above
is dominated by the leftmost point in the left panel of Figure~\ref{f:mbias},
namely A68.  No trend is suggested between X-ray to weak-lensing mass ratio
and fluctuation parameter in the middle panel of Figure~\ref{f:mbias}.

The mass and concentration ($c$) of an individual NFW halo are
anti-correlated.  For example, the covariance for $M^{\rm WL}_{500}$ and
$c_{500}$ is $\sigma_{M}^2=1.25$, $\sigma_{Mc}=-0.44$, and $\sigma_{c}^2=1.20$
for the sample of 12 clusters.  In practical terms, this means that clusters
with higher weak-lensing-based values of $c_{500}/\langle c_{500}\rangle$ will
on average have higher values of $M^{\rm X}/M^{\rm WL}$, in which $\langle
c_{500}\rangle$ is the average of the concentration parameters from Okabe et
al. (2010a) for the sample of 12 clusters, weighted by the inverse square of
the errors.  Most significantly, the uncertainties on $c_{500}/\langle
c_{500}\rangle$ and $M^{\rm X}/M^{\rm WL}$ are correlated, which may induce a
correlation between the quantities themselves.  We therefore plot $M^{\rm
  X}/M^{\rm WL}$ versus $c_{500}/\langle c_{500}\rangle$ in the right panel of
Figure~\ref{f:mbias} --- no obvious relation is seen in this figure, suggesting
that the effect is small and/or the effect is canceled out by other
effects. This result does not change when we use the concentration parameter
$c_{\rm vir}=r_{\rm vir}/r_{\rm s}$.  A much larger sample and more detailed
analysis are required to investigate this issue further.

Results from numerical simulations indicate that the cluster concentration
parameter may be related to the cluster dynamical state (e.g., Neto et
al. 2007). Using the Millennium Simulation, Neto et al.\ pointed out that the
concentrations of out-of-equilibrium halos tend to be lower and have more
scatter compared to their equilibrium counterparts.  This can also be
investigated in the observed $M^{\rm X}/M^{\rm WL}$ versus $c_{500}/\langle
c_{500}\rangle$ plane, however, the current sample is too small.

Finally, we note that the concentration measurements used in this article are
based on weak-lensing constraints (Okabe et al.\ 2010a).  Future papers in
this series will use more precise concentration parameter estimates based on
combined strong- and weak-lensing models of the clusters.  This will allow a
more quantitative comparison between observations and simulations.

\subsection{Gas mass fraction} 
\label{s:fgas}

\begin{figure*}
\begin{center}
\includegraphics[angle=270,width=8cm]{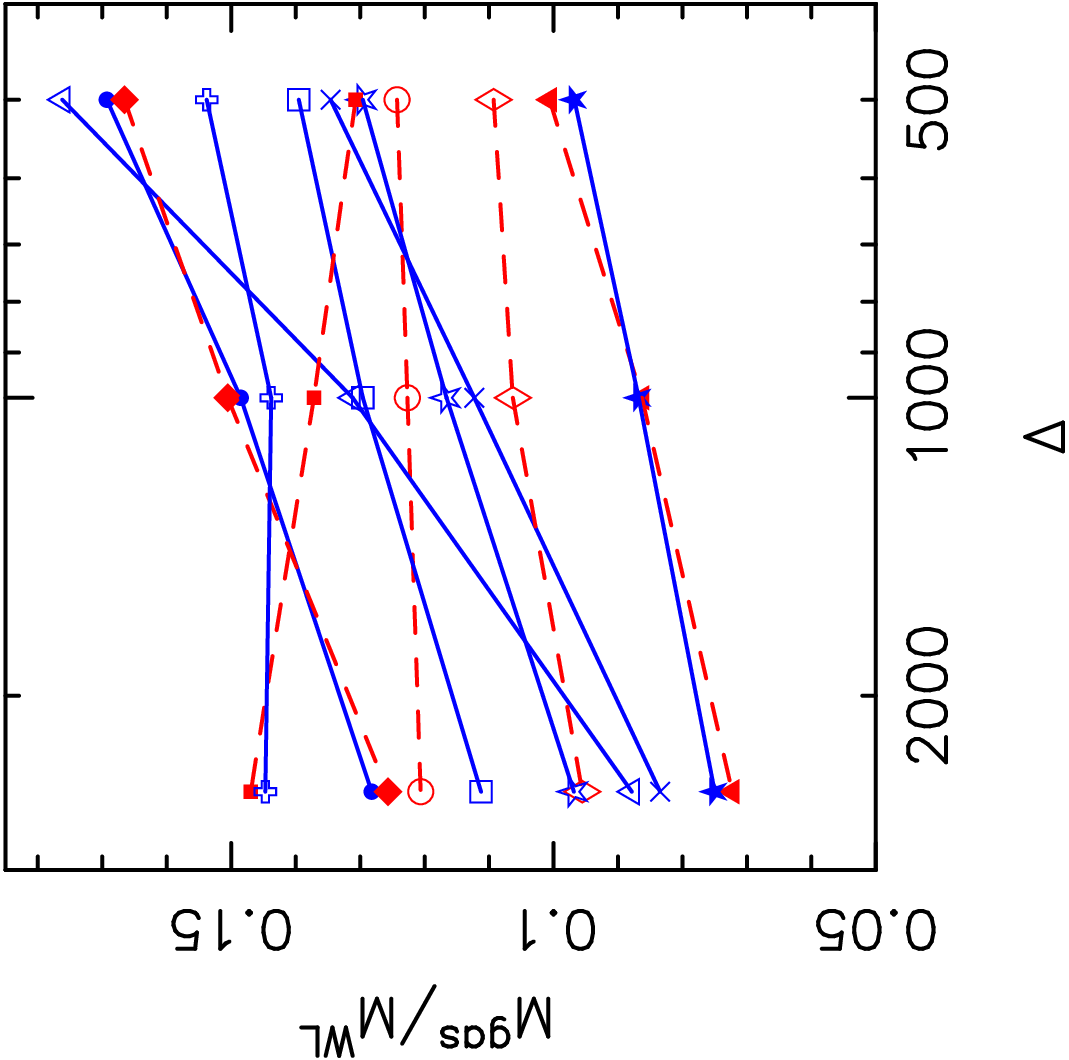}
\includegraphics[angle=270,width=8cm]{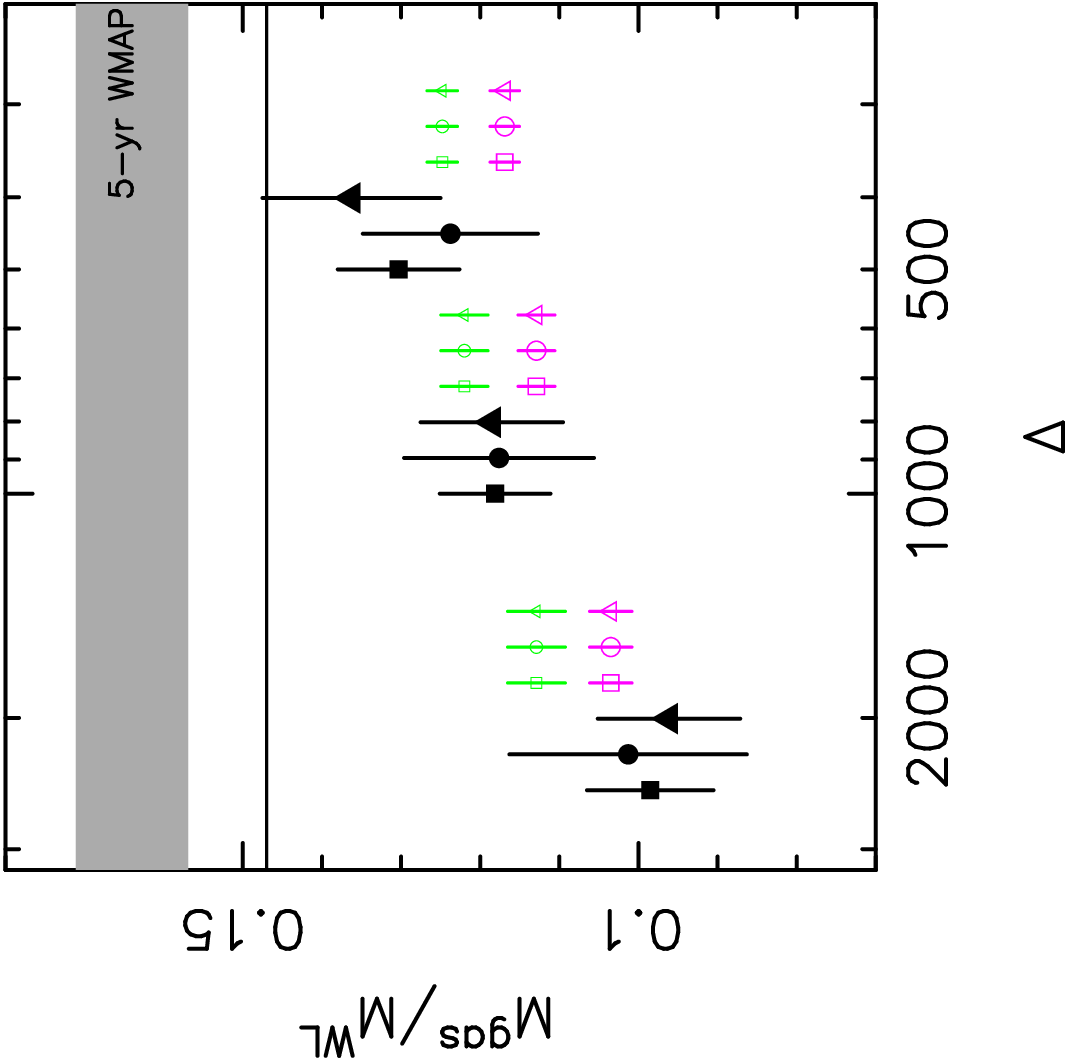}
\end{center}
\caption{{\it Left panel:} gas mass fractions using weak-lensing masses for
  individual clusters. The colors, lines, and symbols are the same as shown in
  the left panel of Figure~\ref{f:mx_to_mwl}. {\it Right panel:} average gas
  mass fractions using weak-lensing masses (filled symbols). Boxes, circles,
  and triangles denote all, undisturbed, and disturbed clusters. The data sets
  are off by 0.048~dex at each density contrast for clarity. The gray
  horizontal band shows the $1\sigma$ range of the cosmic mean baryon fraction
  from \emph{WMAP} five-year data, $\Omega_{\rm b}/\Omega_{\rm
    m}=0.164\pm0.007$, in Komatsu et al. (2009), and the black horizontal line
  corresponds to $0.9 \times$ the mean cosmic baryon fraction. Also shown are
  the gas mass fractions for the sub-samples drawn from the PVge2keV sample
  with mass cuts of $M_{500} \ge 5 \times 10^{14} M_{\odot}$ (small green open
  symbols) and $M_{500} \ge 2.6 \times 10^{14} M_{\odot}$ (big magenta open
  symbols). (A color version of
  this figure is available in the online journal.) }
\label{f:fgas}
\end{figure*}

We define the gas mass fraction as $M^{\rm gas}(\le r^{\rm
  WL}_{\Delta})/M^{\rm WL}(\le r^{\rm WL}_{\Delta})$, and show the gas mass
fractions for individual clusters at $\Delta=2500$, 1000, and 500 in the left
panel of Figure~\ref{f:fgas}. There is a clear trend of increasing gas mass
fraction toward larger cluster-centric radius. A similar trend is also common
in both adiabatic simulations and in simulations with cooling (e.g., Evrard
1990, Lewis et al.  2000, Kravtsov et al. 2005). We also calculate the average
gas mass fractions for the full sample and the undisturbed and disturbed
sub-samples, at $\Delta=2500$, 1000, and 500 --- right panel of Figure~\ref{f:fgas}
and Table~\ref{t:fgas}. The average gas mass fraction shows no dependence on
cluster morphology, and approaches the cosmic mean baryon fraction
$\Omega_{\rm b}/\Omega_{\rm m}=0.164\pm0.007$ (Komatsu et al.  2009) at large
cluster-centric radius.  At $\Delta\sim500$, the average gas mass fractions
stand at $90\%$ of the cosmic mean value, in good agreement with simulations
(e.g., Nagai et al.\ 2007).

Our gas mass fractions are also in good agreement with Umetsu et al.'s (2009)
joint Subaru weak-lensing and AMiBA Sunyaev--Zel'dovich effect (SZE) analysis
of four clusters.  Umetsu et al.\ obtained $\langle f_{\rm
  gas,2500}\rangle=0.105\pm0.015\pm0.012$ and $\langle f_{\rm
  gas,500}\rangle=0.126\pm0.019\pm0.016$, where the first error is the
statistical error, and the second one is the standard error from the average
due to cluster-to-cluster variance. However, the gas mass fraction obtained by
Mahdavi et al.\ (2008) using weak-lensing and X-ray data is slightly higher
than our results: $\langle f_{\rm gas,2500}\rangle=0.119\pm0.006$.  A detailed
understanding of this $\sim2\sigma$ discrepancy awaits analysis of our
complete volume-limited sample in a future article.

Several X-ray-only studies have advocated gas mass fractions as a potential
probe of the dark energy equation of state parameter $w$ (Vikhlinin et al.\
2006; Allen et al.\ 2008).  We therefore investigate how our
weak-lensing-based gas mass fractions compare with X-ray-only gas mass
fractions, defining the latter as $M^{\rm gas}(\le r^{\rm X}_{\Delta})/M^{\rm
  X}(\le r^{\rm X}_{\Delta})$ --- i.e., the X-ray-only gas mass fractions are
measured at radii determined from the X-ray data derived hydrostatic mass
distribution.  We calculated the average X-ray-only gas mass fractions at
$\Delta=2500$, 1000, and 500 for the full sample and the undisturbed/disturbed
sub-samples (Table~\ref{t:fgasx}).  We found that the X-ray-only gas mass
fractions are statistically indistinguishable from weak-lensing-based gas mass
fractions at all over-densities considered and for all three samples, except
for disturbed clusters at $\Delta=500$, for which the X-ray-only value appears
to be biased low.  The most significant result in the context of cluster
cosmology is the very close agreement at $\Delta=2500$ between the respective
gas mass fraction measurements for undisturbed and disturbed clusters,
regardless of whether the cluster mass measurement is based on X-ray or
lensing data.

\begin{figure*}
\begin{center}
\includegraphics[angle=270,width=8cm]{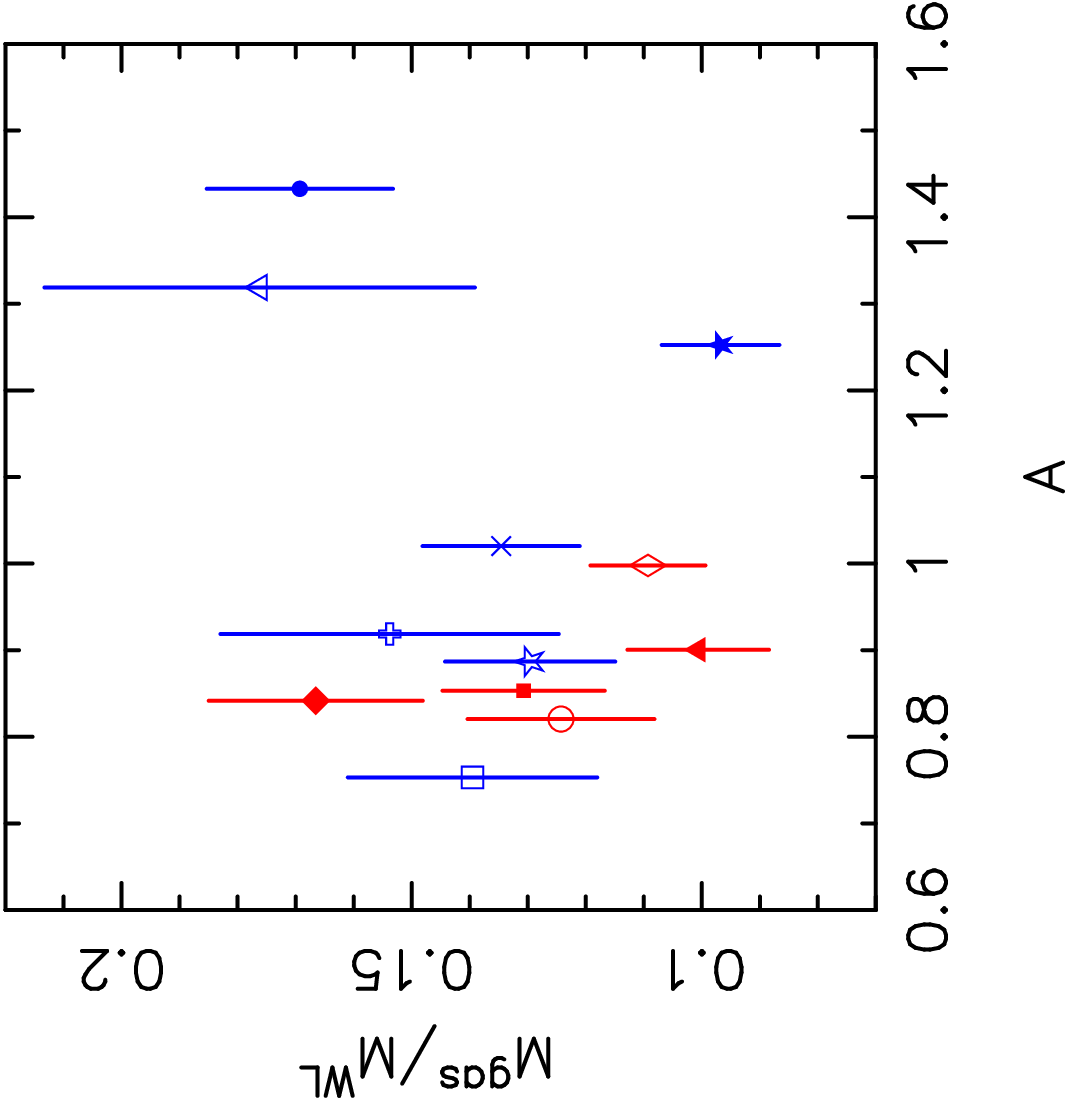}
\includegraphics[angle=270,width=8cm]{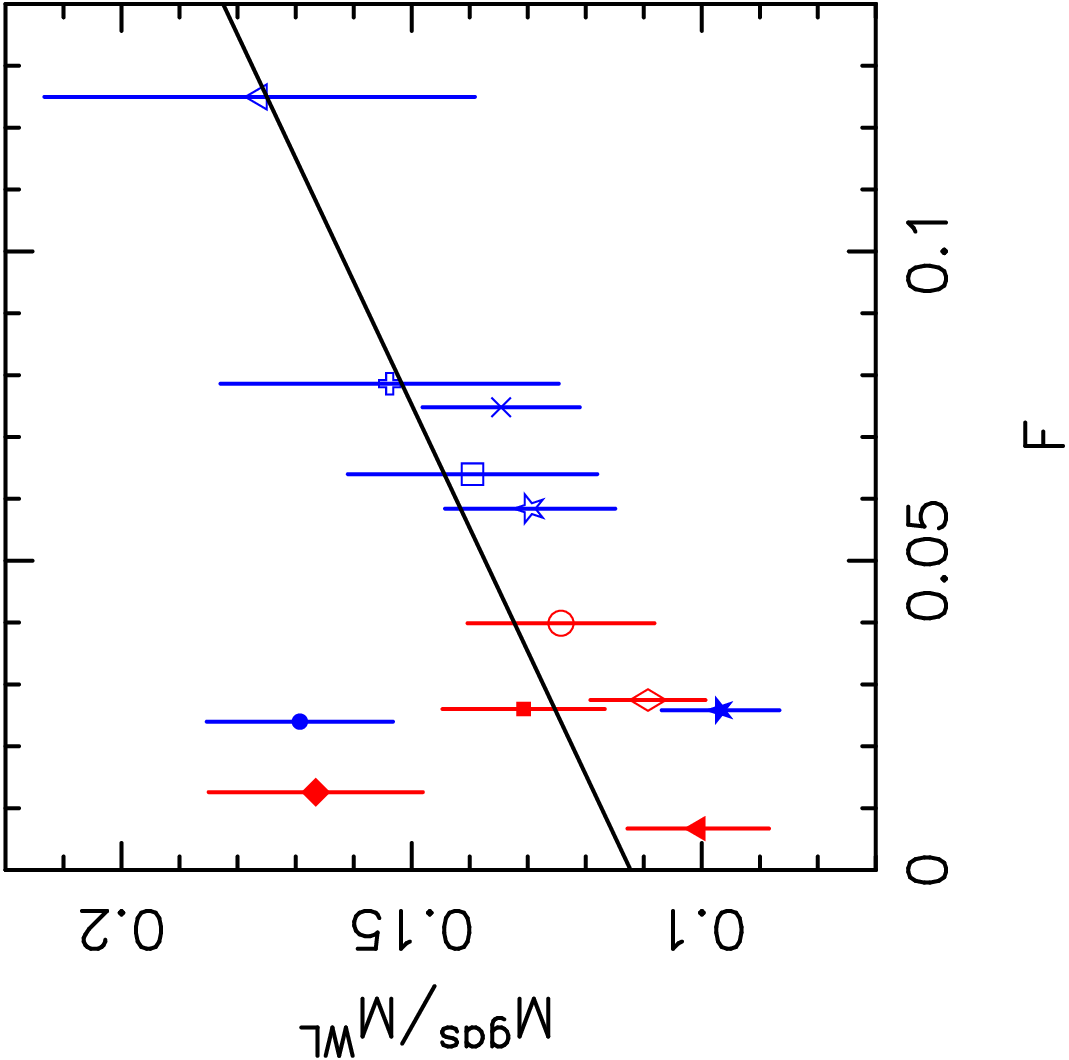}
\end{center}
\caption{Gas mass to weak-lensing mass ratio at the radius with $\Delta=500$
  vs. asymmetry parameter (left panel) and fluctuation parameter (right
  panel). The best fit with equal weighting of each data point is shown in the
  right panel. The colors and symbols are the same as shown in the left panel
  of Figure~\ref{f:mx_to_mwl}. (A color version of this figure is available in
  the online journal.)}
\label{f:mgbias}
\end{figure*}

We examine this apparent morphological independence of gas mass fraction
measurements by looking at the possible dependence of gas mass fraction
measurements of individual clusters on the asymmetry and fluctuation
parameters.  We plot gas mass fraction versus asymmetry and fluctuation
parameters in Figure~\ref{f:mgbias}.  The former resembles a scatter plot as
might be expected based on the results above.  However, undisturbed and
disturbed clusters both follow the same trend (right panel of
Figure~\ref{f:mgbias}) of increasing gas mass fraction with fluctuation
parameter.  This may indicate a direct connection between the gas mass
fraction and substructure fraction, i.e., the fraction of cluster mass that
resides in massive sub-halos (substructures) within the cluster halo.  Large
substructure fractions are generally attributed to recent infall of galaxy
groups/clusters into more massive clusters --- see Richard et al.\ (2010) for
more details.  In a future article, we will investigate the relationship
between weak-lensing-based substructure fractions and cluster X-ray
morphology, including the fluctuation parameter.  For now, we fit a straight
line to all 12 data points in the right panel of Figure~\ref{f:mgbias},
obtaining: $M^{\rm gas}/M^{\rm WL}= (0.112 \pm 0.011) +(0.502 \pm 0.197)\cdot
F$ at $\Delta=500$.

Finally, we compare the gas mass fraction with the weak-lensing mass.  In this
current narrow mass range, there is no obvious dependence on the weak-lensing
mass. However, X-ray studies for samples with broad mass ranges have shown
that there is a strong mass dependence (e.g., Pratt et al. 2009).  This will
be discussed later in Section~\ref{s:dis}.

\section{Discussion}
\label{s:dis}

In this section, we compare our results with recent numerical simulations of
clusters.  There is a broad consensus among numerical studies that total
cluster masses derived from X-ray data underestimate the true cluster masses
in both undisturbed and disturbed clusters for variety of reasons, most
notably because of the deviation from H.E. and extra pressure support (e.g.,
turbulence) beside the thermal gas (e.g., Evrard 1990; Lewis et al. 2000;
Miniati et al. 2001; Miniati 2003; Rasia et al.  2006; Nagai et al. 2007; Pfrommer et al. 2007; PV08; Jeltema et
al. 2008; Lau et al. 2009; Meneghetti et al. 2010).  An obvious difference
between numerical and observational studies is that the true cluster mass is
known in the former, and not in the latter. One solution is to construct fake
weak-lensing data from the simulated cluster data (e.g., Meneghetti et
al. 2008, 2010).  However, large samples of such ``observed'' theoretical
clusters are not yet available.  We therefore rely on the insensitivity of the
weak-lensing mass measurements to cluster thermodynamics to assume that our
weak-lensing masses are, on average, well matched to ``true'' masses of the
numerical studies.  We therefore treat weak-lensing-based mass measurements as
the ``truth'' in the comparisons described in this section.

\subsection{Summary of recent simulated cluster samples}

Before describing the comparison with simulations in the next section, we
first summarize the key features of the simulated cluster samples against
which we compare our results.  We discuss in turn the samples of Nagai et al.\
(2007), Lau et al.\ (2009), PV08, and Jeltema et al.\ (2008).

Nagai et al. (2007) studied 16 simulated clusters with $T>2$~keV in the mass
range of $M_{500}\sim(0.3 - 9)\times10^{14}M_\odot h^{-1}$. They derived
hydrostatic masses by analyzing mock \emph{Chandra} data (three projections per
cluster) following the method in Vikhlinin et al. (2006), which is similar to
our reduction method (Zhang et al.\ 2008).  We adopt Nagai et al.'s
morphological classifications --- their 48 simulated X-ray cluster observations
comprise 21 undisturbed (``relaxed'' in their terminology) clusters and 27
disturbed (``unrelaxed'') clusters.

Lau et al. (2009) studied the same simulated clusters as Nagai et al., but
estimated hydrostatic masses directly using the three-dimensional gas
profiles.  Importantly, Lau et al.\ used mass weighted temperatures ($T_{\rm
  MW}$), in contrast to Nagai et al.'s temperatures that were reconstructed
from mock observations.  Lau et al.\ suggest that this difference explains the
different average mass biases reported in the two papers. The reconstructed
temperature profiles used by Nagai et al.\ are not spectroscopic ones, because
they are derived by weighting the contribution of temperature components along
the line of sight to correct for the spectroscopic bias. Nevertheless, they
are derived from spectroscopic data and sensitive to the local multiphase
structure of the gas.  In contrast to PVge2keV (see below) and Nagai et al.,
Lau et al.\ define a cluster as relaxed/undisturbed only if it appears so in
all three projections, which gives 6 relaxed/undisturbed clusters and 10
unrelaxed/disturbed clusters in their sample. The different fractions of
undisturbed and disturbed clusters found by Nagai et al.\ and Lau et al.\
(based on the same simulated sample) arise from the fact that the perceived
dynamical state can depend on viewing angle when using simple diagnostics.

PV08 analyzed $100$ simulated clusters (three projections per cluster) with $T >
2$~keV ($8.2\times 10^{13}M_\odot h^{-1}\simlt M_{200}\simlt 1.2\times
10^{15}M_\odot h^{-1}$) --- this is to date the largest, complete
volume-limited sample of simulated clusters for which the hydrostatic mass
biases have been investigated in detail. They applied different techniques to
measure the X-ray hydrostatic mass in order to disentangle biases of different
origin. Here, we compare our results with their average biases derived
adopting an extended $\beta$-model to fit the gas density radial profile to
estimate the hydrostatic mass, which is similar to the procedure in Vikhlinin
et al.\ (2006), Nagai et al.\ (2007), and Zhang et al.\ (2008), and
three-dimensional mass-weighted temperature profiles ($T_{\rm MW}$) or
spectroscopic-like temperature ($T_{\rm SL}$; Mazzotta et al 2004).

In order to improve the coverage at high mass end, we use an extended version
of the PV08 sample, which is extracted from a larger cosmological simulation.
The sample is constructed and analyzed exactly as the PV08 sample and yields
results fully consistent with the latter. The new sample (PVge2keV sample,
hereafter) comprises $\sim 120$ clusters (3 projections per cluster) with $T >
2$~keV.  Average mass biases derived for the $T_{\rm MW}$ case are fully
consistent with those derived using directly the three-dimensional gas
profiles. The PVge2keV sample of 360 simulated X-ray cluster observations were
divided into 180 undisturbed clusters and 180 disturbed clusters using their
mock X-ray data; this matches the roughly 50--50 split between disturbed and
undisturbed clusters in the observed sample.

Jeltema et al. (2008) analyzed 61 simulated clusters giving a 16\% bias of the
hydrostatic mass estimate on average at $\Delta=500$. However, their analysis
is restricted to just $\Delta=500$, and their X-ray hydrostatic masses are
derived using the true three-dimensional gas profiles only. We therefore
chose not to include a detailed comparison with their results. Nevertheless,
notice that their results are fully consistent with those presented in this
paper.

In summary, the different hydrostatic mass reconstruction methods, and in
particular temperature definitions, adopted in the simulations discussed above
imply that it is sensible to compare our observational results with (1)
PVge2keV ($T_{\rm MW}$ case) and Lau et al. (2009), and (2) PVge2keV ($T_{\rm
  SL}$ case) and Nagai et al. (2007), with an emphasis on the latter to match
our use of observed spectral temperatures.  We note that the simulated
clusters extend to lower temperatures ($T\gs2$keV) than the observed sample
($T\ge5$keV).  However, the numerical studies have shown that the dependence of
hydrostatic mass bias on cluster temperature (mass) is negligible. The way to
define the cluster dynamical state in PV08 and Nagai et al. (2007) should be
more consistent with the one in our observations. However, for our comparison
there is no point to correct the difference because the errors are tiny for
the samples in simulations compared to the errors for the observational
sample.  Finally, we note that the cosmological parameters used in the
simulations are slightly different from those used in this article; the impact
of these small differences on our results/discussion is negligible.

\subsection{Comparing observational mass estimates to simulations}

We overplot the X-ray hydrostatic to weak-lensing mass ratios from the
simulations in the right panel of Figure~\ref{f:mx_to_mwl}.  We find that
there is a general agreement between observations and simulations in which the
mass ratios are consistent with unity at small radii, e.g., at $\Delta=2500$,
and the agreement deteriorates at larger radii.  Most strikingly, the
simulations agree well with our result that the mass ratio for undisturbed
clusters declines gently to larger radii.  We estimate that a firm detection
of the discrepancy between simulations and observations for undisturbed
clusters, would require a sample of $\sim60$ clusters.

For disturbed clusters, the simulations --- particularly those using the
spectroscopic-like temperatures --- show a significant discrepancy between the
X-ray hydrostatic and true masses at large radii, while no discrepancy is
detected observationally between the X-ray hydrostatic and weak-lensing
masses.  The hydrostatic mass bias seen in simulations, when spectroscopic
temperatures are used in the derivation of hydrostatic masses, might be
exacerbated by over-cooling.  Even when small-scale cold clumps of gas are
masked in the mock analysis, over-cooling can lead to underestimated spectral
temperatures because of the presence of a diffuse cold gas component.  This
may be the cause of the disagreement between the observed sample of 12
clusters and the simulated samples at large radii.

\subsection{Comparing observational gas mass fractions with simulations}

Within a given density contrast, the gas mass fraction tends to increase with the
increasing cluster mass (e.g., Vikhlinin et al. 2006, Gonzalez et al. 2007,
Gastaldello et al. 2007, Pratt et al. 2009), with the trend being shallower at
large radii. This behavior can be explained by the scale dependency introduced
by cooling (Bryan 2000), which regulates the amount of cool gas present in
clusters. Numerical simulations can also reproduce the observed behavior of
$f_{\rm gas}$ in clusters (Valdarnini 2003, Kravtsov et al. 2005, Ettori et
al.2006, Kay et al. 2007), and strongly support the cooling model.

Due to the strong mass dependence of gas mass fractions seen in clusters, a
proper comparison between sample averages requires that the two samples
approximately cover the same mass range. This is accomplished by extracting a
sub-sample of PVge2keV clusters that all have masses above the lowest mass of
observed clusters: $M_{500} \ge 2.6 \cdot 10^{14} M_{\odot}$. This sub-sample
contains $N_{\rm sub}=68$ clusters which are subsequently divided into $34$
undisturbed and $34$ disturbed clusters based on the cluster dynamical state
as explained in PV08. The average gas mass fractions for all 68 simulated
clusters and the undisturbed and disturbed simulated clusters are compared to
the observed gas mass fractions in Figure~\ref{f:fgas}. To illustrate the
strong mass dependence of gas mass fractions, we also show in
Figure~\ref{f:fgas} average gas mass fractions for a more restrictive
sub-sample of 32 PVge2keV clusters with $M_{500}\ge5\cdot10^{14}M_\odot$.

The average gas mass fractions of simulated clusters agree with those of
observed clusters in the sense that they do not show any dependence on cluster
morphology. The agreement between simulations and observations is most
striking at $\Delta=2500$, where the observed and simulated gas mass fractions
agree within the uncertainties when the mass range of the respective samples
is matched.  Overall there is a good agreement within the error bars, taking
into account that the trend with mass is strong and that the mass function of
the two samples (from simulations and observations) is not exactly the same.

\subsection{X-ray and lensing masses in simulated clusters}

As discussed at the beginning of Section~\ref{s:dis}, large samples of mock
weak-lensing observations of simulated clusters are not yet available.  The
largest sample to date is that of Meneghetti et al. (2010), who studied three
clusters, with three orthogonal projections per cluster, to generate a total set
of nine mock observations.  Their X-ray hydrostatic masses are typically biased
low by 5\%-20\% due to the lack of H.E. in their simulated clusters, which is in
agreement with our results.  They also found the gas mass to be well
reconstructed within the region where the X-ray surface brightness profile is
extracted.  Their gas mass measurements are independent of the dynamical state
of the cluster, with average deviations of $1\%\pm3$\% at $\Delta=2500$ and
$7\%\pm4$\% at $200<\Delta<500$.  Although Meneghetti et al.'s sample is small,
their results agree well with our observational results that the gas mass to
weak-lensing mass ratios are independent of the cluster dynamical state, and
supports our finding that the radius with $\Delta=2500$ appears to be the most
robust radius at which to use cluster gas mass fractions for probing
cosmological parameters.

\section{Summary and outlook} 
\label{s:con}

We have presented a comparison of X-ray hydrostatic and weak-lensing mass
estimates for 12 clusters at $z\simeq0.2$ for which high-quality
\emph{XMM-Newton} and Subaru/Suprime-CAM data are available within the LoCuSS.
Our main results are as follows:
 
\begin{itemize}

\item For the full sample, we obtain $1-M^{\rm X}/M^{\rm WL}=0.01 \pm 0.07$ at
  an over-density of $\Delta=500$.  We also sub-divided the sample into
  undisturbed and disturbed sub-samples based on quantitative X-ray
  morphologies using asymmetry and fluctuation parameters. We obtained
  $1-M^{\rm X}/M^{\rm WL}=0.09 \pm 0.06$ and $-0.06 \pm 0.12$ for the
  undisturbed and disturbed clusters, respectively.

\item The scatter around the average X-ray hydrostatic to weak-lensing mass
  ratio for undisturbed clusters is half that of the disturbed clusters due to
  the lower intrinsic variance among the population of undisturbed clusters.

\item For disturbed clusters, it is unclear whether the X-ray hydrostatic mass
  is consistent with the weak-lensing mass, due to large scatter. In addition
  to non-thermal pressure support, there may be a competing effect associated
  with adiabatic compression and/or shock heating of the intracluster gas
  which leads to overestimate of X-ray H.E. masses for disturbed clusters. The
  most prominent example of this in our sample is the famous merging cluster
  A1914.

\item Despite the modest statistical significance of the mass discrepancy in
  the undisturbed clusters, we detect a clear trend of improving agreement
  between $M^{\rm X}$ and $M^{\rm WL}$ as a function of increasing
  over-density, $M^{\rm X}/M^{\rm WL}=(0.908 \pm 0.004)+(0.187 \pm 0.010)
  \cdot \log_{10} (\Delta/500)$.

\item There is a general agreement between the X-ray hydrostatic to
  weak-lensing mass ratio in observed and simulated clusters in which the mass
  ratios are both consistent with unity at small radii (i.e., at
  $\Delta=2500$), with the agreement deteriorating at larger radii, i.e., out
  to $\Delta=500$.  This deterioration is dominated by disturbed clusters,
  with the undisturbed simulated clusters reproducing well the observed gentle
  decline in $M^{\rm X}/M^{\rm WL}$ as a function of increasing
  cluster-centric radius.
 
\item The weak-lensing mass-based cumulative gas mass fraction increases with
  radius, but still lies below the cosmic baryon fraction at the largest
  cluster-centric radii probed (i.e., $\Delta=500$). An important finding is
  the absence of dependence of the gas mass fraction on cluster dynamical
  state (i.e., X-ray morphology). The X-ray-only gas mass fractions are also
  consistent with the weak-lensing mass-based gas mass fractions at
  $\Delta=2500$, supporting the proposal to use this measurement as a probe of
  the dark energy equation of state parameter $w$.

\end{itemize}

In summary, our results demonstrate that \emph{XMM-Newton} and Subaru are a
powerful combination for calibrating systematic uncertainties in cluster mass
measurements and suggest an encouraging convergence between X-ray, lensing,
and numerical studies of cluster mass.  Nevertheless, our observational sample
remains very small at just 12 clusters.  Our detailed results are therefore
vulnerable to inclusion/exclusion of extreme clusters, as highlighted by our
discussion of A1914.  Robust constraints on systematic uncertainties in
cluster mass measurement therefore await a thorough investigation of a large
and complete volume-limited sample of clusters, such as that planned within
LoCuSS.  On the theoretical side, it is important to move as rapidly as
possible toward large samples of mock weak-lensing and X-ray observations of
simulated clusters, such as those recently pioneered by Meneghetti et al.\
(2008, 2010).  With both of these observational and theoretical data sets in
hand, a detailed and statistically robust comparison will be possible, helping
to calibrate future cluster cosmology experiments.

\acknowledgments We acknowledge support from KICP in Chicago for hospitality,
and thank our LoCuSS collaborators, especially Masahiro Takada and Keiichi
Umetsu, for helpful comments on the manuscript. Y.Y.Z. thanks Massimo
Meneghetti and Gabriel Pratt for useful discussion. Y.Y.Z. acknowledges
support by the DFG through Emmy Noether Research Grant RE~1462/2, through
Schwerpunkt Program 1177, and through project B6 ``Gravitational Lensing and
X-ray Emission by Non-Linear Structures'' of Transregional Collaborative
Research Centre TRR 33 – “The Dark Universe”, and support by the German BMBF
through the Verbundforschung under grant 50\,OR\,0601. This work is supported
by a Grant-in-Aid for the COE Program ``Exploring New Science by Bridging
Particle-Matter Hierarchy'' and G-COE Program ``Weaving Science Web beyond
Particle-Matter Hierarchy'' in Tohoku University, funded by the Ministry of
Education, Science, Sports and Culture of Japan. This work is, in part,
supported by a Grant-in-Aid for Science Research in a Priority Area "Probing
the Dark Energy through an Extremely Wide and Deep Survey with Subaru
Telescope" (18072001) from the Ministry of Education, Culture, Sports,
Science, and Technology of Japan. N.O. is, in part, supported by a
Grant-in-Aid from the Ministry of Education, Culture, Sports, Science, and
Technology of Japan (20740099). A.F. acknowledges support from BMBF/DLR under
grant 50\,OR\,0207 and MPG, and was partially supported by a NASA grant
NNX08AX46G to UMBC. G.P.S. acknowledges support from the Royal Society and
STFC. D.P.M. acknowledges support provided by NASA through Hubble Fellowship
grant \#HF-51259.01 awarded by the Space Telescope Science Institute, which is
operated by the Association of Universities for Research in Astronomy, Inc.,
for NASA, under contract NAS 5-26555.

\clearpage

\begin{table}
\begin{center}
  \caption[]{Cluster centers from weak-lensing and X-ray analysis\\ }
\begin{tabular}{l|rrrrrrc}
  \hline
  Cluster Name & \multicolumn{2}{c}{X-ray Center [J2000]} & \multicolumn{2}{c}{Weak-lensing Center [J2000]}  & \multicolumn{2}{c}{Offset} & X-ray Morphology \\ 
  & R.A. & Decl. & R.A. & Decl. & (arcmin) & $r_{2500}$ & \\
  \hline
  A68            &  00 37 06.2 &  09 09 28.7 &  00 37 06.9 &  09 09 24.5 & 0.18 & 0.10 &Disturbed\\ 
  A115 (south)   &  00 56 00.3 &  26 20 32.5 &  00 56 00.3 &  26 20 32.5 & 0.00 & 0.00 &Disturbed\\ 
  A209           &  01 31 52.6 & -13 36 35.5 &  01 31 52.5 & -13 36 40.5 & 0.08 & 0.04 &Undisturbed\\ 
  A267           &  01 52 42.0 &  01 00 41.2 &  01 52 41.9 &  01 00 25.7 & 0.26 & 0.14 &Disturbed\\ 
  A383           &  02 48 03.3 & -03 31 43.6 &  02 48 03.4 & -03 31 44.7 & 0.03 & 0.01 &Undisturbed\\ 
  A1835          &  14 01 01.9 &  02 52 35.5 &  14 01 02.1 &  02 52 42.8 & 0.13 & 0.06 &Undisturbed\\ 
  A1914          &  14 26 00.9 &  37 49 38.8 &  14 25 56.7 &  37 48 59.2 & 1.22 & 0.44 &Disturbed\\ 
  Z7160          &  14 57 15.2 &  22 20 31.2 &  14 57 15.1 &  22 20 35.3 & 0.07 & 0.05 &Disturbed\\ 
  A2261          &  17 22 26.9 &  32 07 47.4 &  17 22 27.2 &  32 07 57.1 & 0.34 & 0.13 &Disturbed\\ 
  A2390          &  21 53 37.1 &  17 41 46.4 &  21 53 36.8 &  17 41 43.3 & 0.08 & 0.03 &Undisturbed\\ 
  RXCJ2129.6+0005&  21 29 39.8 &  00 05 18.5 &  21 29 40.0 &  00 05 21.8 & 0.07 & 0.04 &Undisturbed\\ 
  RXCJ2337.6+0016&  23 37 37.8 &  00 16 15.5 &  23 37 39.7 &  00 16 17.0 & 0.48 & 0.25 &Disturbed\\ 
  \hline
\end{tabular}
\label{t:cen}
\end{center}
\hspace*{0.3cm}{\footnotesize Note. Within the sample, 
  A1914 and RXCJ2337.6+0016 show extreme
  offsets between the X-ray and weak-lensing determined cluster central
  positions. Therefore, the X-ray analysis of the two clusters was revised
  using the weak-lensing determined cluster centers. }
\end{table}

\begin{table}
\begin{center}
  \caption[]{Cluster mass measurements$^{a}$\\ }
\begin{tabular}{l|ccccccccc}
  \hline
  Cluster Name & \multicolumn{3}{c}{$r^{\rm WL}_{2500}$} & \multicolumn{3}{c}{$r^{\rm WL}_{1000}$} & \multicolumn{3}{c}{$r^{\rm WL}_{500}$} \\ 
  & $M^{\rm WL}$ & $M^{\rm X}$ & $M^{\rm gas}$ & $M^{\rm WL}$ & $M^{\rm X}$ & $M^{\rm gas}$ &$M^{\rm WL}$ & $M^{\rm X}$ & $M^{\rm gas}$ \\
  \hline
  A68       &$ 1.42^{+0.59}_{-0.62} $&$  1.91\pm 0.57 $&$  0.158\pm 0.007  $&$ 2.78^{+0.79}_{-0.78}  $&$  3.56\pm 1.06 $&$  0.360\pm 0.026  $&$ 4.15^{+1.21}_{-1.06}   $&$  5.50\pm 1.65  $&$  0.579\pm 0.057  $\\ 
A115 (south)&$ 1.22^{+0.65}_{-0.67} $&$  0.68\pm 0.12 $&$  0.107\pm 0.005  $&$ 2.51^{+0.93}_{-0.91}  $&$  1.98\pm 0.50 $&$  0.329\pm 0.017  $&$ 3.85^{+1.61}_{-1.32}   $&$  3.84\pm 0.99  $&$  0.679\pm 0.037  $\\ 
  A209      &$ 2.18^{+0.46}_{-0.46} $&$  1.95\pm 0.55 $&$  0.208\pm 0.014  $&$ 5.24^{+0.74}_{-0.72}  $&$  4.41\pm 1.40 $&$  0.557\pm 0.050  $&$ 8.81^{+1.31}_{-1.20}   $&$  6.57\pm 1.95  $&$  0.963\pm 0.101  $\\ 
  A267      &$ 1.43^{+0.25}_{-0.25} $&$  1.66\pm 0.45 $&$  0.139\pm 0.007  $&$ 2.41^{+0.42}_{-0.40}  $&$  2.58\pm 0.86 $&$  0.281\pm 0.021  $&$ 3.29^{+0.68}_{-0.61}   $&$  3.75\pm 1.16  $&$  0.426\pm 0.039  $\\ 
  A383      &$ 1.76^{+0.21}_{-0.20} $&$  1.61\pm 0.48 $&$  0.127\pm 0.008  $&$ 2.64^{+0.43}_{-0.40}  $&$  2.42\pm 0.72 $&$  0.227\pm 0.022  $&$ 3.38^{+0.71}_{-0.62}   $&$  3.23\pm 0.95  $&$  0.340\pm 0.041  $\\ 
  A1835     &$ 2.88^{+0.57}_{-0.58} $&$  2.98\pm 0.89 $&$  0.423\pm 0.023  $&$ 6.15^{+0.95}_{-0.90}  $&$  5.66\pm 1.68 $&$  0.844\pm 0.080  $&$ 9.65^{+1.70}_{-1.51}   $&$  8.59\pm 2.50  $&$  1.261\pm 0.154 $\\ 
  A1914     &$ 2.09^{+0.31}_{-0.31} $&$  2.78\pm 0.76 $&$  0.268\pm 0.015  $&$ 3.35^{+0.50}_{-0.47}  $&$  4.36\pm 1.22 $&$  0.497\pm 0.038  $&$ 4.46^{+0.75}_{-0.69}   $&$  7.69\pm 2.24  $&$  0.754\pm 0.066 $\\ 
  Z7160     &$ 0.89^{+0.38}_{-0.42} $&$  1.17\pm 0.35 $&$  0.129\pm 0.004  $&$ 1.75^{+0.58}_{-0.55}  $&$  1.80\pm 0.52 $&$  0.251\pm 0.014  $&$ 2.61^{+0.97}_{-0.82}   $&$  2.46\pm 0.72  $&$  0.401\pm 0.033 $\\ 
  A2261     &$ 3.55^{+0.44}_{-0.44} $&$  2.77\pm 0.75 $&$  0.266\pm 0.026  $&$ 5.95^{+0.74}_{-0.71}  $&$  4.54\pm 1.20 $&$  0.516\pm 0.065  $&$ 8.12^{+1.23}_{-1.12}   $&$  6.24\pm 1.65  $&$  0.786\pm 0.114 $\\ 
  A2390     &$ 3.14^{+0.44}_{-0.43} $&$  3.92\pm 1.15 $&$  0.395\pm 0.031  $&$ 5.22^{+0.82}_{-0.77}  $&$  6.02\pm 1.94 $&$  0.785\pm 0.083  $&$ 7.09^{+1.29}_{-1.17}   $&$  7.48\pm 2.22  $&$  1.180\pm 0.150 $\\ 
  RXCJ2129.6&$ 1.38^{+0.53}_{-0.54} $&$  1.75\pm 0.52 $&$  0.166\pm 0.009  $&$ 2.97^{+0.69}_{-0.71}  $&$  3.07\pm 0.93 $&$  0.364\pm 0.031  $&$ 4.68^{+1.09}_{-0.97}   $&$  4.46\pm 1.29  $&$  0.581\pm 0.065  $\\ 
  RXCJ2337.6&$ 2.42^{+0.36}_{-0.37} $&$  1.97\pm 0.44 $&$  0.202\pm 0.015  $&$ 3.72^{+0.50}_{-0.47}  $&$  3.16\pm 0.68 $&$  0.418\pm 0.045  $&$ 4.85^{+0.75}_{-0.70}   $&$  4.45\pm 0.97  $&$  0.652\pm 0.084  $\\ 
  \hline
\end{tabular}
\label{t:mass}
\end{center}
  \hspace*{0.3cm}{\footnotesize Note. $^{a}$~The values are in units of $10^{14}$
    solar mass. }
\end{table}

\begin{table}
\begin{center}
  \caption[]{Comparison of X-ray and weak-lensing mass estimates\\}
\begin{tabular}{l|ccc}
  \hline
  Sample          & \multicolumn{3}{c}{$M^{\rm X}/M^{\rm WL}$}\\ 
                  &  $r^{\rm WL}_{2500}$ & $r^{\rm WL}_{1000}$  & $r^{\rm WL}_{500}$  \\ 
  \hline
  All             & $1.01\pm0.07$ &  $0.97\pm0.05$ &  $0.99\pm0.07$ \\ 
  Undisturbed     & $1.04\pm0.08$ & $0.96\pm0.05$ & $0.91\pm0.06$ \\ 
  Disturbed       & $0.98\pm0.12$ & $0.97\pm0.09$ & $1.06\pm0.12$ \\ 
  All-A1914       & $0.97\pm0.07$ & $0.94\pm0.05$ & $0.94\pm0.05$ \\ 
  Disturbed-A1914 &$0.92\pm0.11$ & $0.91\pm0.07$ & $0.97\pm0.08$  \\ 
  \hline
\end{tabular}
\label{t:mratio}
\end{center}
\end{table}

\begin{table}
\begin{center}
  \caption[]{Results of Monte Carlo simulations$^{a}$\\}
\begin{tabular}{l|ccc}
  \hline
Sample        & \multicolumn{3}{c}{$M^{\rm X}/M^{\rm WL}$}\\
              &  $r^{\rm WL}_{2500}$ & $r^{\rm WL}_{1000}$  & $r^{\rm WL}_{500}$   \\ 
  \hline
  All         & $1.03\pm0.09 \pm 0.13$ & $0.94\pm0.06 \pm 0.05$ & $0.95\pm0.07 \pm 0.08$ \\ 
  Undisturbed & $1.02\pm0.05 \pm 0.02$ & $0.94\pm0.04 \pm 0.01$ & $0.90\pm0.04 \pm 0.02$ \\ 
  Disturbed   & $1.05\pm0.11 \pm 0.22$ & $0.94\pm0.07 \pm 0.08$ & $1.02\pm0.08 \pm 0.11$ \\ 
  \hline
\end{tabular}
\label{t:pdf}
\end{center}
\hspace*{0.3cm}{\footnotesize Note. $^{a}$~The values quoted for each comparison 
  are the mean, standard error, and real variance based on 500 Monte Carlo 
  simulations described in Section~\ref{s:biasobs}.} 
\end{table}

\begin{table}
\begin{center}
  \caption[]{Weak-lensing-based gas mass fractions\\}
\begin{tabular}{l|ccc}
\hline 
Sample      & \multicolumn{3}{c}{$M^{\rm gas}/M^{\rm WL}$}\\ 
  & $r^{\rm WL}_{2500}$ & $r^{\rm WL}_{1000}$  & $r^{\rm WL}_{500}$  \\ 
\hline
All         & $0.099\pm0.008$ & $0.118\pm0.007$ & $0.130\pm0.008$ \\
Undisturbed & $0.101\pm0.015$ & $0.118\pm0.012$ & $0.124\pm0.011$ \\
Disturbed   & $0.096\pm0.009$ & $0.119\pm0.009$ & $0.136\pm0.011$ \\
\hline
\end{tabular}\label{t:fgas}
\end{center}
\hspace*{0.3cm}{\footnotesize } 
\end{table}

\begin{table}
\begin{center}
  \caption[]{X-ray-only gas mass fractions\\}
\begin{tabular}{l|ccc}
\hline 
Sample      & \multicolumn{3}{c}{$M^{\rm gas}/M^{\rm X}$}\\
            &  $r^{\rm X}_{2500}$ & $r^{\rm X}_{1000}$  & $r^{\rm X}_{500}$ \\ 
\hline
All         & $0.101\pm0.010$ & $0.118\pm0.015$ & $0.131\pm0.015$ \\
Undisturbed & $0.100\pm0.017$ & $0.122\pm0.028$ & $0.124\pm0.023$ \\
Disturbed   & $0.101\pm0.018$ & $0.115\pm0.024$ & $0.110\pm0.026$ \\
\hline
\end{tabular}\label{t:fgasx}
\end{center}
\hspace*{0.3cm}{\footnotesize } 
\end{table}

\end{document}